\documentclass[letterpaper]{article} 
\usepackage{aaai2026}  
\usepackage{times}  
\usepackage{helvet}  
\usepackage{courier}  
\usepackage[hyphens]{url}  
\usepackage{graphicx} 
\usepackage{natbib}  
\usepackage{caption} 
\usepackage{algorithm}
\usepackage{algorithmic}

\usepackage{newfloat}
\usepackage{listings}
\DeclareCaptionStyle{ruled}{labelfont=normalfont,labelsep=colon,strut=off} 
\floatstyle{ruled}
\newfloat{listing}{tb}{lst}{}
\floatname{listing}{Listing}
\usepackage{amsmath}
\usepackage{amssymb}
\usepackage{amsthm}
\usepackage[capitalize,noabbrev]{cleveref}
\usepackage{graphicx}
\usepackage{stmaryrd}
\usepackage{nicefrac}
\usepackage{algorithm}
\usepackage{algorithmic}
\usepackage{caption}
\usepackage{subfigure}
\usepackage{url}
\usepackage{multirow}

\theoremstyle{plain}
\newtheorem{theorem}{Theorem}[section]

\newtheorem{lemma}[theorem]{Lemma}

\theoremstyle{definition}
\newtheorem{definition}[theorem]{Definition}

\theoremstyle{remark}

\title{Shapley Value Approximation Based on k-Additive Games}
\author {
    Guilherme Dean Pelegrina\equalcontrib \textsuperscript{\rm 1},
    Patrick Kolpaczki\equalcontrib \textsuperscript{\rm 2,3},
    Eyke H\"ullermeier \textsuperscript{\rm 2,3,4}
}
\affiliations {
    \textsuperscript{\rm 1}Engineering School - Mackenzie Presbyterian University\\
    \textsuperscript{\rm 2}LMU Munich\\
    \textsuperscript{\rm 3}Munich Center for Machine Learning (MCML)\\
    \textsuperscript{\rm 4}German Research Center for Artificial Intelligence (DFKI, DSA)\\
    guilherme.pelegrina@mackenzie.br, patrick.kolpaczki@lmu.de, eyke@lmu.de
}

\begin{document}

\maketitle

\begin{abstract}
    The Shapley value is the prevalent solution for fair division problems in which a payout is to be divided among multiple agents.
    By adopting a game-theoretic view, the idea of fair division and the Shapley value can also be used in machine learning to quantify the individual contribution of features or data points to the performance of a predictive model.
    Despite its popularity and axiomatic justification, the Shapley value suffers from a computational complexity that scales exponentially with the number of entities involved, and hence requires approximation methods for its reliable estimation.
    We propose \emph{SVA}$k_{\text{ADD}}$, a novel approximation method that fits a $k$-additive surrogate game.
    By taking advantage of $k$-additivity, we are able to elicit the exact Shapley values of the surrogate game and then use these values as estimates for the original fair division problem. 
    The efficacy of our method is evaluated empirically and compared to competing methods.
\end{abstract}

\begin{links}
    \link{Code}{https://github.com/GuilhermePelegrina/Shapley_Value_Approximation_Based_on_k-Additive_Games.git}
\end{links}

\section{Introduction}
\label{sec:intr}

The complexity of machine learning models experienced a rapid and certainly significant increase over the last decade.
This development comes with an ever-rising burden to understand a model's decision-making, reaching a point beyond human comprehension.
Meanwhile, societal and political influences led to a growing demand for trustworthy AI~\citep{Li2023}.
The field of Explainable AI (XAI) emerges to counteract these consequences, aiming to bring back understanding to the human user.
Among various explanation types~\citep{Molnar2021}, post-hoc additive explanations convince with an intuitive appeal: an observed numerical effect caused by the behavior of the black box model is divided among participating entities.
Additive feature explanations decompose a predicted value \cite{Lundberg.2017} or generalization performance \citep{Covert.2020} among the involved features, enabling feature importance scores.
Beyond explainability, this allows in feature engineering to conduct feature selection by removing features with irrelevant or even harmful contributions \citep{Cohen.2005, Marcilio2020}.

Treating this decomposition as a fair division problem opens the door to game theory which views the features as cooperating agents, forming groups called coalitions to achieve a task and collect a common reward that is to be shared.
Such scenarios are captured by the widely applicable notion of cooperative games, modeling the agents as a set of players $N$ and assuming that a real-valued worth $\nu(A)$ can be assigned to each coalition $A \subseteq N$ by a value function $\nu$.
Among multiple propositions the Shapley value \citep{Shapley.1953} prevailed as the most favored solution to the fair division problem.
It assigns to each player a share of the collective benefit, more precisely a weighted average of all its marginal contributions, i.e., the increase in collective benefit a player causes when joining a coalition.
Its popularity is rooted in the fact that it is provably the only solution to fulfill certain desirable axioms \citep{Shapley.1953} which arguably capture a widespread understanding of fairness.
For example, in supply chain cooperation~\citep{Fiestras-Janeiro2011}, the cost reduction when joining a coalition may be shared among companies based on the Shapley value.
The greater a company's marginal contributions to the cost reduction, the greater its received payoff.

The applicability of the Shapley value exceeds economics as its utility has been recognized within various disciplines.
Most prominently, it has found its way into the field of machine learning, especially as a model-agnostic approach, quantifying the importance of entities such as features or datapoints (see \citep{Rozemberczki.2022} for an overview).
Adopting the game-theoretic view, these entities are understood as players which cause a certain numerical outcome.
Shaping the measure of a coalition's worth adequately is pivotal to the informativeness of the importance scores obtained by the Shapley values.
For example, considering a model's generalization performance on a test dataset restricted to the feature subset given by a coalition yields global feature importance scores \citep{Pfannschmidt.2016, Covert.2020}.
Conversely, local feature attribution scores are obtained by splitting the model's prediction value for a fixed datapoint \citep{Lundberg.2017}.
The Shapley value is not limited to provide additive explanations since it has also been proposed to perform data valuation \citep{Ghorbani.2019}, feature selection \citep{Cohen.2007} by removing features with low relevance towards the model's performance \citep{Pelegrina2024}, ensemble construction \citep{Rozemberczki.2021}, and the pruning of neural networks \citep{Ghorbani.2020}.

Further practical applications include its usage to quantify each feature's impact in predicting the risk degree in managing industrial machine maintenance \cite{Nimmy2023},
\citet{Pelegrina2023b} apply it to evaluate the influence of each electrode on the quality of recovered fetal electrocardiograms, and \citet{Brusa2023} measure the features' importance towards machinery fault detection.
Worth mentioning, each application requires an appropriate modeling in terms of player set and value function in order to obtain meaningful scores.

The uniqueness of the Shapley value comes at a price that poses an inherent drawback to practitioners: its computation scales exponentially with the number of players taking part in the cooperative game.
Consequently, it becomes quickly infeasible for increasing feature numbers or even a few datapoints, especially when complex models are in use whose evaluation is highly resource consuming.
As a viable remedy, it is common practice to approximate the Shapley value while providing reliably precise estimates is crucial to obtain meaningful importance scores.
On this background, the recent interest in assigning scores to features, datapoints, or even model components,
has fueled the research on approximation algorithms, leading to a diverse landscape of approaches (see \citep{Chen.2023} for an overview).

\paragraph{Contribution.}
We propose with \emph{SVA}$k_{\text{ADD}}$ (Shapley Value Approximation under $k$-additivity) a novel approximation method for the Shapley value based on the concept of $k$-additive games whose structure elicits a denser parameterizable value function.
Fitting a $k$-additive surrogate game to randomly sampled coalition-value pairs comes with a twofold benefit.
First, it reduces flexibility, promising faster convergence and second, the Shapley values of the $k$-additive surrogate game are obtained immediately from its representation.
In summary, our contributions are:

\begin{itemize}
    \item \emph{SVA}$k_{\text{ADD}}$ fits a $k$-additive surrogate game to sampled coalitions, mimicking the given game by a simpler structure with parameterizable degree of freedom while maintaining low representation error.
    The surrogate game's own Shapley values are obtained immediately and yield precise estimates for the given game if the representation exhibits a good fit.
    
    \item \emph{SVA}$k_{\text{ADD}}$ does not require any structural properties of the value function.
    Thus, it is domain-independent and can be applied to any cooperative game oblivious to what players and payoffs represent.
    Specifically in the field of explainability, it is model-agnostic and can approximate local as well as global explanations.

    \item We prove the theoretical soundness of \emph{SVA}$k_{\text{ADD}}$ by showing analytically that its underlying optimization problem yields the Shapley value.
   
     \item We empirically compare \emph{SVA}$k_{\text{ADD}}$ to competitive baselines at the hand of various explanation tasks, and shed light onto the best fitting degree of $k$-additivity.
\end{itemize}

\section{Related Work}
\label{sec:related_work}

The problem of approximating the Shapley value, tackled by various communities, lead to a multitude of approaches to overcome its complexity.
First to mention among the class of methods that can handle arbitrary games, without further assumptions on the value function, are those which construct mean estimates via random sampling.
Fittingly, the Shapley value of a player can be interpreted as its expected marginal contribution to a specific probability distribution over coalitions.
\citet{Castro.2009} propose with \emph{ApproShapley} the sampling of permutations from which marginal contributions are extracted.
Further works employ stratification by coalition size \citep{Maleki.2013,Castro.2017,vanCampen.2018, Okhrati.2020, Zhang.2023}, or utilize reproducing kernel Hilbert spaces \citep{Mitchell.2022}.
Departing from marginal contributions, \emph{Stratified SVARM} \citep{Kolpaczki.2024a} splits the Shapley value into multiple means of coalition values being further refined by \emph{Adaptive SVARM} \citep{Kolpaczki.2024b}.
Guided by a different representation of the Shapley value, \emph{KernelSHAP} \citep{Lundberg.2017} solves an approximated weighted least squares problem, to which the Shapley value is its solution.
\citet{Fumagalli.2023} prove its variant \emph{Unbiased KernelSHAP} \cite{Covert.2021} to be equivalent to importance sampling of single coalitions.
Joining this family, \citet{Pelegrina.2023} propose $k_{ADD}$-SHAP, which formulates the surrogate model assuming a $k$-additive game\footnote{
    Note that $k_{ADD}$-SHAP is limited to local explanations.
    In contrast, our proposed method \emph{SVA}$k_{\text{ADD}}$ differs by its applicability to any formulation of a cooperative game.
    Moreover, in the context of explainable AI, it is capable of providing global explanations.
}.
It locally adopts the Choquet integral as the interpretable model, whose parameters have a straightforward connection with the Shapley value.
Similar to us, \citet{Yan.2020} apply surrogate games of parameterizable structure that sum up unanimity games to calculate Shapley values, however, under the assumption that the given game possesses this structure from which we refrain.

On the contrary, tailoring the approximation to a specific application by leveraging structural properties promises faster converging estimates.
In data valuation, including knowledge of how datapoints tend to contribute to a learning algorithm's performance resulted in multiple tailored methods \cite{Ghorbani.2019, Jia.2019a, Jia.2019b}.
In similar fashion \citet{Liben-Nowell.2012} leverage supermodularity in cooperative games.
Even further, value functions of certain parameterized shapes facilitate closed-form polynomial solutions of the Shapley value w.r.t.\ the number of players.
Examples include the voting game \citep{Bilbao.2000} and the minimum cost spanning tree games \citep{Granot.2002} used in operations~research.
\section{The Shapley Value and $k$-Additivity}
\label{sec:theory}

We formally introduce cooperative games and the Shapley value in \cref{subsec:games_shapley}.
Next, we present in \cref{subsec:k_additivity} the concept of $k$-additivity, constituting the core of our approach.

\subsection{Cooperative Games and the Shapley Value}
\label{subsec:games_shapley}

A cooperative game is formally described by $n$ players, captured by the set $N = \{1,\ldots,n\}$, and an associated payoff function $\nu : \mathcal{P}(N) \to \mathbb{R}$, where $\mathcal{P}(N)$ represents the power set of $N$.
This simple but expressive formalism may for example represent a shipment coordination where companies form a coalition in order to save costs when delivering their products.
In this case, the companies can be modeled as players
and $\nu(A)$ represents the benefit achieved by the group of companies $A \subseteq N$.
Clearly, $\nu(N)$ is the total benefit when all companies (players) form the grand coalition $N$. Commonly, one normalizes the game by defining $\nu(\emptyset) = 0$, i.e., the worth of the empty set.
However, in 
explainability, $\nu(\emptyset)$ may take nonzero values, e.g., with no features available one may obtain a classification accuracy of 50\%. 
In this case, one can normalize $\nu$ by simply subtracting the worth of the empty set from all game payoffs, i.e., $\nu'(A) \leftarrow \nu(A) - \nu(\emptyset)$ for all $A \subseteq N$.

A central question arising from a cooperative game is how to fairly share the worth $\nu(N)$ of the grand coalition $N$ among all participating players. 
The Shapley value \citep{Shapley.1953} emerges as the prevalent solution concept since it uniquely satisfies axioms that intuitively capture fairness \citep{Shapley.1953}.
Given the game $(N, \nu)$, the Shapley value of each player $i$ is defined as
\begin{equation}
    \label{eq:shapley}
    \phi_i = \sum\limits_{A \subseteq N \setminus \{i\}} \frac{\left(n - \left| A \right| - 1 \right)! \left| A \right|!}{n!} [\nu(A \cup \{i\}) - \nu(A)] \, ,
\end{equation}
where $\left| A \right|$ represents the cardinality of coalition $A$.
It can be interpreted as a player's weighted average of marginal contributions to the payoff.
Among the fulfilled axioms such as null player, symmetry, and additivity (see~\citep{Young1985} for more details and other properties), in explainability the most useful is efficiency.
It demands that the sum of all players' Shapley values is equal to the difference between $\nu(N)$ and $\nu(\emptyset)$.
Formally, efficiency means
$\sum\nolimits_{i=1}^n \phi_i = \nu(N) - \nu(\emptyset)$.
Or, in the game theory framework where $\nu(\emptyset) = 0$, one obtains $\sum_{i=1}^n \phi_i = \nu(N)$.
In explainability, efficiency can be used to decompose a measure of interest among the set of features.
As a result, one can interpret the importance of each feature to that measure.

Unfortunately, satisfying the desired axioms in the form of the Shapley value comes at a price.
According to \cref{eq:shapley}, the calculation requires the evaluation of all $2^n$ coalitions within the exponentially growing power set of $N$.
In fact, the exact computation of the Shapley value is known to be NP-hard \citep{Deng.1994}.
Hence, its exact computation does not only become practically infeasible for growing player numbers but it is also of interest that the evaluation of only a few coalitions suffices to retrieve precise estimates.
For instance, a model has to be costly re-trained and re-evaluated on a test dataset for each coalition if one is interested in the features' impact on the generalization performance.
Therefore, a common goal is to approximate all Shapley values $\phi = (\phi_1,\ldots,\phi_n)$ of a given game $(N,\nu)$ by observing only a subset of evaluated coalitions $ \mathcal{M} \subseteq \mathcal{P}(N)$.
We denote the size of $\mathcal{M}$ by $T \in \mathbb{N}$ and refer to it as the available budget representing the number of samples an approximation algorithm is allowed to draw.
The mean squared error (MSE) serves as a popular measure to quantify the quality of the obtained estimates $\hat\phi = (\hat\phi_1,\ldots,\hat\phi_n)$ and is to be minimized:
$\frac{1}{n} \sum\nolimits_{i=1}^n \mathbb{E} [ ( \hat\phi_i - \phi_i )^2 ]$,
where the expectation is taken w.r.t.\ the (potential) randomness of the approximation strategy.

\subsection{Interaction Indices and $k$-Additivity}
\label{subsec:k_additivity}

The underlying idea of measuring the impact (or share) of a single player $i$ by means of its marginal contributions finds its natural extension to sets of players $S$ in the Shapley interaction index~\citep{Murofushi1993,Grabisch1997a} by generalizing from marginal contributions to discrete derivatives.
For any $S \subseteq N$ its Shapley interaction $I(S)$ is given by
\begin{equation}
\label{eq:inter_ind}
     I(S) = \sum_{A \subseteq N\backslash S} w_{A,S} \left( \sum_{A' \subseteq S} \left(-1\right)^{\left| S \right| - |A'|}\nu(A \cup A') \right)
\end{equation}
with weights $w_{A,S} = \frac{\left(n-\left|A\right|-\left|S\right|\right)!\left|A\right|!}{\left(n-\left|S\right|+1\right)!}$.
For convenience, we will write $I_i := I(\{i\})$ and $I_{i,j} := I(\{i,j\})$.
Instead of individual importance, $I(S)$ indicates the synergy between players in $S$.
Although this interpretation is not straightforward for coalitions of three or more entities, it has a clear meaning for pairs.
For two players $i$ and $j$, the Shapley interaction index $I_{i,j}$ quantifies how the presence of $i$ impacts the marginal contributions of $j$ and vice versa.
Especially~in explainable AI, where players represent features, it can be interpreted as follows: (i) if $ I_{i,j} < 0$, there is a negative interaction (redundant effect) between features $i,j$; (ii) if $I_{i,j} > 0$, there is a positive interaction (complementary effect) between features $i,j$; (iii) if $I_{i,j} = 0$, there is no interaction between $i,j$ (independence) on average.

\noindent
Note that the Shapley interaction index reduces to the Shapley value for a singleton, i.e., $I_i = \phi_i$.
Moreover, there is a linear relation between the interactions and the game payoffs~\citep{Grabisch1997a}. Indeed, from the interactions one may easily retrieve the game payoffs by the following expression:
\begin{equation}
\label{eq:iitomu_s}
\nu(A) = \sum_{B \subseteq N} \gamma^{\left|B \right|}_{\left| A \cap B \right|}I(B) \, ,
\end{equation}
where $\gamma^{\left|B \right|}_{\left| A \cap B \right|}$ is defined as follows with the Bernoulli numbers starting at $\eta_0=1$:
\begin{equation*}
\gamma_{r}^{s} = \sum_{l=0}^{r}\binom{r}{l}\eta_{s-l}
\hspace{0.5cm} \text{and} \hspace{0.5cm}
\eta_{r} = -\sum_{l=0}^{r-1}\frac{\eta_{l}}{r-l+1}\binom{r}{l} \, .
\end{equation*}
This linear transformation recovers any coalition value $\nu(A)$ by using the Shapley interactions of all $2^n$ coalitions, thus including the Shapley values.
Therefore, $2^n$ parameters are to be defined if the whole game is to be expressed by Shapley interactions.
However, in some situations one may assume that interactions only exist for coalitions up to $k$ many players. This assumption leads to the concept known as $k$-additive games.
A $k$-additive game is such that $I(S) = 0$ for all $S$ with $\left| S \right| > k$.
Obviously, this restricts the flexibility of the game but depending on $k$, this may significantly decrease the number of parameters to be defined such that for low $k$ it increases only polynomially with the number of players. 
For instance, in $2$-additive and $3$-additive games, there are only $n(n+1)/2$, and $n(n^2+5)/6$ respectively, many interactions indices as the remaining parameters are equal to zero.
One may argue that within Shapley-based feature explanations, the neglect of higher order interactions, by setting them to zero per default, comes naturally.
For instance, \cite{Bordt.2023} show that these interactions barely exist in the context of post-hoc local explanations.
\section{$k$-Additive Approximation Approach}
\label{sec:proposal}

\begin{figure}[t]
\centering
\includegraphics[width=0.99\columnwidth]{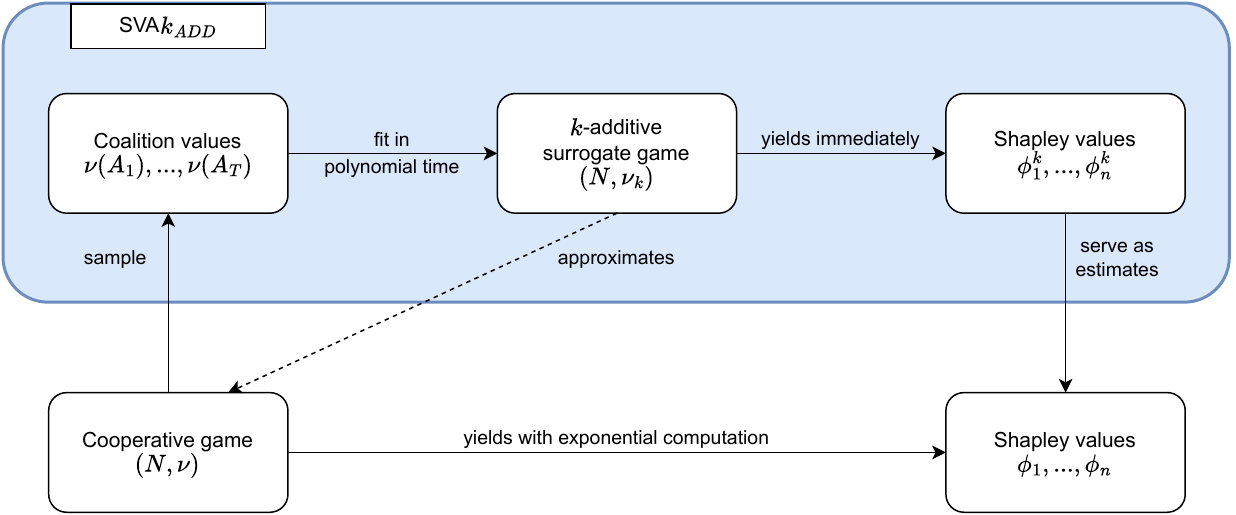}
\caption{
The sampled coalition values $\nu(A_1),\ldots,\nu(A_T)$ from the game $(N,\nu)$ are used to fit a $k$-additive~\mbox{surrogate} game $(N,\nu_k)$ in polynomial time w.r.t.\ $n$.
The Shapley values $\phi_1^k,\ldots,\phi_n^k$ of $(N,\nu_k)$ are obtained immediately~from~its $k$-additive representation.
As $\nu_k$ approximates $\nu$, these serve as estimates of the true Shapley values $\phi_1,\ldots,\phi_n$ of $(N,\nu)$.
}
\label{fig:kAdd}
\end{figure}

In this section, we present our method \emph{SVA}$k_{\text{ADD}}$ to approximate Shapley values.
It builds upon the idea of adjusting a $k$-additive surrogate game $(N,\nu_k)$ to randomly sampled and evaluated coalitions.
Having fit the surrogate game to represent the observed coalition values with minimal error, its own Shapley values $\phi^k$ can be interpreted as estimates $\hat\phi$ for $\phi$ of $(N,\nu)$ since the fitting promises $\nu_k$ to be close to $\nu$.
See \cref{fig:kAdd} for an illustration of the approach.
The framework of fitting a surrogate game encompasses other methods such as \emph{KernelSHAP} \cite{Lundberg.2017} and $k_{ADD}$-SHAP \cite{Pelegrina.2023}.
See Appendix~\ref{app:comparison} for a conceptual comparison with our proposal.

\subsection{The $k$-Additive Optimization Problem}
\label{subsec:optimization}

We leverage the representation of $\nu_k$ by means of interactions as given in \cref{eq:iitomu_s}.
In particular, since $\nu_k$ is supposed to be $k$-additive, we specify $\nu_k$ as a linear transformation of interactions $I^k(B)$ for all $B \subseteq N$ of size $|B| \leq k$, allowing us to truncate interactions of higher order than $k$: 
\begin{equation}
    \nu_k(A) = \sum_{B \subseteq N, |B| \leq k} \gamma^{\left|B \right|}_{\left| A \cap B \right|} I^k(B) \, .
\end{equation}
Within this representation, the Shapley values $\phi^k$ of the resulting game $(N,\nu_k)$ are obtained immediately by the interactions $I_i^k = \phi_i^k$, which will serve as estimates for the Shapley values $\phi$ of the game $(N,\nu)$, i.e.\ $I_i^k \approx \phi_i$.
The $k$-additive representation of $\nu_k$ comes with the advantage that the number of parameters $I^k(B)$ needed to define the surrogate game is reduced (as several parameters are set to zero).
The drawback of this strategy is the reduction in flexibility left to model the observed game $(N,\nu)$ according to the obtained evaluations.
However, we can still model interactions for coalitions up to $k$ players.
Empirically, works in the literature~\citep{Grabisch2002,Grabisch2006,Pelegrina2020,Pelegrina.2023} have been using $2$-additive or even $3$-additive games and obtained satisfactory results for modeling interactions.
Our goal is to fit $\nu_k$ as closely as possible to $\nu$ and therefore minimize the deviation of $\nu_k$ from $\nu$ captured by
\begin{equation}
    \sum_{A \in \mathcal{P}(N) \setminus \{\emptyset, N\}} w_A \left( \nu(A) - \nu_k(A) \right)^2 \, ,
\end{equation}
where $w_A$ is an importance weight associated to each coalition $A$.
We are eager to meet the desirable efficiency axiom such that the difference between and $\nu(N)$ and $\nu(\emptyset)$ is decomposed among the players within our approximated values $\phi^k$.
This is ensured by imposing the constraint $\nu(N) - \nu(\emptyset) = \nu_k(N) - \nu_k(\emptyset)$.
Hence, we arrive at the following optimization problem.
\begin{definition} \label{def:optimization}
    Given a cooperative game $(N, \nu)$, a degree of $k$-additivity $k \in \mathbb{N}$ with $k \leq n$, and weights $w_A \in \mathbb{R}$ associated with each coalition $A \subseteq N$, the $k$-additive optimization problem is given by the following constrained weighted least square optimization problem:
    \begin{equation*}
        \begin{array}{rl}
            \displaystyle\min_{I^k} & \hspace{-0.3cm} \sum\limits_{A \in \mathcal{P}(N) \setminus \left\{\emptyset, N \right\}} \hspace{-0.1cm} w_A \left( \hspace{-0.05cm} \nu(A) - \hspace{-0.1cm}\sum\limits_{B \subseteq N, |B| \leq k} \hspace{-0.1cm} \gamma^{\left|B \right|}_{\left| A \cap B \right|} I^k(B) \hspace{-0.05cm} \right)^2 \\
            \text{s.t.} & \nu(N) - \nu(\emptyset) \ = \sum\limits_{B \subseteq N, |B| \leq k} \left( \gamma^{\left|B \right|}_{\left| B \right|} - \gamma^{\left| B \right|}_0 \right) I^k(B)
        \end{array}
    \end{equation*}
\end{definition}
\noindent
Solving the $k$-additive optimization is at the core of our approach.
In the remainder we describe how to overcome two key challenges.
First, we address in \cref{subsec:theory} how to choose the weights $w_A$ such that $\phi^k$ comes close to $\phi$.
Second, as the objective function sums up over exponential many coalitions, we present in \cref{subsec:sampling} our algorithm \emph{SVA}$k_{\text{ADD}}$ that constructs an approximative objective function by sampling coalitions and adding their error terms.

\subsection{Theoretical Soundness Through Choice of Weights}
\label{subsec:theory}

Seeking precise estimates $\phi^k \approx \phi$, one may even raise the question if it is feasible to retrieve the exact Shapley values $\phi$ from the solution $I^k$ and how the weights $w_A$ have to be set to achieve this.
We analytically derive the correct weights and positively answer this question.
\begin{theorem} \label{the:solution}
    The solution to the $k$-additive optimization problem of any cooperative game $(N,\nu)$ for the cases of $k=1$, $k=2$, and $k=3$ with weights $w_A^* = \binom{n-2}{|A|-1}^{-1}$ yields the Shapley value, i.e.\
    \begin{equation*}
        I_i^k = \phi_i \, .
    \end{equation*}
\end{theorem}
\noindent
See Appendix~\ref{app:analysis} for the proof of \cref{the:solution}.
Note that the weights coincide with those derived by \cite{Charnes.1988} used in \citep{Lundberg.2017} for a different optimization problem.
The result implies that after observing all coalitions of the cooperative game $(N,\nu)$ our approach yields the exact Shapley values with no approximation error.
We interpret this as evidence for the soundness and theoretical foundation of our method.
Moreover, since the result holds irregardless of the shape of $\nu$, the game can even highly deviate from being $k$-additive and our estimates will still converge to $\phi$.
Hence, $k$-additivity is not an assumption that our method requires but rather a tool to be leveraged.
We conjecture that \cref{the:solution} holds also true for arbitrary degrees of $k$-additivity and leave the proof for future work due to the analytical challenges.
Worth mentioning is that the hardness of incorporating Shapley interactions of higher degree into weighted least squares optimizations has been acknowledged by \cite{Fumagalli.2024}.

\subsection{Approximating the $k$-Additive Optimization Problem via Sampling}
\label{subsec:sampling}

Computing the solution to the $k$-additive optimization problem (see \cref{def:optimization}) is practically infeasible since the objective compromises exponential many error terms w.r.t.\ $n$.
As a remedy we follow the same strategy as adopted in~\citep{Lundberg.2017,Pelegrina.2023} and approximate the objective function by sampling coalitions without replacement.
Let $\mathcal{M} = \{A_1,\ldots,A_T\}$ be the set of sampled coalitions with $A_i \neq A_j$ for all $i \neq j$ and the sequence $\nu_{\mathcal{M}} = (\nu(A_1),\ldots,\nu(A_T))$ representing its evaluated coalition values.
Thus, we solve the following optimization problem after sampling:
\begin{equation}
\label{eq:opt_kadd}
\begin{array}{rl}
\displaystyle\min_{I^k} & \hspace{-0.3cm} \sum\limits_{A \in \mathcal{M}\backslash \left\{\emptyset, N \right\}} w_{A} \left( \nu(A) - \sum\limits_{B \subseteq N, |B| \leq k} \gamma^{\left|B \right|}_{\left| A \cap B \right|}I^k(B) \right)^2 \\
 \text{s.t.} & \nu(N) - \nu(\emptyset) = \sum\limits_{B \subseteq N, |B| \leq k} \left( \gamma^{\left|B \right|}_{\left| B \right|} - \gamma^{\left|B \right|}_0 \right) I^k(B)
\end{array}
\end{equation}
To ensure the efficiency constraint, we force the evaluation of $\nu(\emptyset)$ and $\nu(N)$.
Each coalition $A \in \mathcal{P}(N) \setminus \{\emptyset, N\}$ is drawn according to an initial probability distribution $p$ defined by
$p_A = \frac{w_A^*}{\sum\nolimits_{B \in \mathcal{P}(N) \setminus \{\emptyset, N\}} w_B^*}$ (see Appendix~\ref{app:sampling} for a practical realization of the sampling).
After drawing a coalition $A$, we set $p_A$ to zero and normalize the remaining probabilities.
This procedure is repeated until $\left| \mathcal{M} \right| = T$. 
Algorithm~\ref{alg:proposal} presents the pseudo-code of \emph{SVA}$k_{\text{ADD}}$.
The algorithm requires the game $(N,\nu)$, the additivity degree $k$, and the budget $T$.
It starts by evaluating $\nu(\emptyset)$ and $\nu(N)$.
Thereafter, based on the (normalized) distribution $p$, it samples $T-2$ coalitions from $\mathcal{P}(N) \setminus \{\emptyset, N\}$, evaluates each, and extends $\mathcal{M}$ as well as $\nu_{\mathcal{M}}$.
Finally, it solves the optimization problem in \cref{eq:opt_kadd} with weights $w_A^*$ given by \cref{the:solution} (see Appendix~\ref{app:solution} for an analytical solution).
The extracted Shapley values $\phi^k$ of $\nu_k$ are returned as estimates $\hat\phi$ for the Shapley values $\phi$ of $(N,\nu)$.

\begin{algorithm}[t]
    \caption{\emph{SVA}$k_{\text{ADD}}$}
    \label{alg:proposal}
		\begin{algorithmic}[1]
			\STATE \textbf{Input:} $(N,\nu)$, $k$, $T$		
            \STATE $\mathcal{M} \leftarrow \{\emptyset, N\}$
            \STATE $\nu_{\mathcal{M}} \leftarrow (\nu(\emptyset), \nu(N))$
            \WHILE{$\left| \mathcal{M} \right| < T$}
                \STATE Sample a coalition $A \in \mathcal{P}(N) \setminus \{\emptyset, N\}$ from normalized distribution $p$
                \STATE $\mathcal{M} \leftarrow \mathcal{M} \cup \{A\}$
                \STATE $\nu_{\mathcal{M}} \leftarrow (\nu_{\mathcal{M}},\nu(A))$
                \STATE $p_A \leftarrow 0$
            \ENDWHILE
            \STATE $(I^k(B))_{B \subseteq N : |B| \leq k} \leftarrow \textsc{\texttt{Solve}}(\mathcal{M},\nu_{\mathcal{M}},k)$ 
            \STATE \textbf{Output:} $I_1^k,\ldots,I_n^k$
    \end{algorithmic}
\end{algorithm}

We would like to emphasize that \cref{the:solution} does not make a statement about $I_i^k$ during sampling when not all coalitions are observed.
To the best of our knowledge, there exists no approximation guarantee for methods that estimate the Shapley value by means of a weighted least squares optimization problem.
The difficulty of obtaining a theoretical result is further elaborated by \cite{Covert.2021}.

In addition, we apply a modified version of border sampling utilized by \cite{,Fumagalli.2023,Kolpaczki.2024c,Lundberg.2017} that demonstrates empirical improvements.
Since the coalition sizes at the ends of the spectrum close to $0$ or $n$ comprise relatively few coalitions, we first deterministically collect all coalitions of size $1$,$2$,$n-2$, and $n-1$ and then continue with the random sampling of coalitions with cardinality between $3$ and $n-3$.

\begin{figure*}[t]
\centering
\begin{minipage}[c]{0.28\textwidth}
    \centering
    \includegraphics[width=0.99\textwidth]{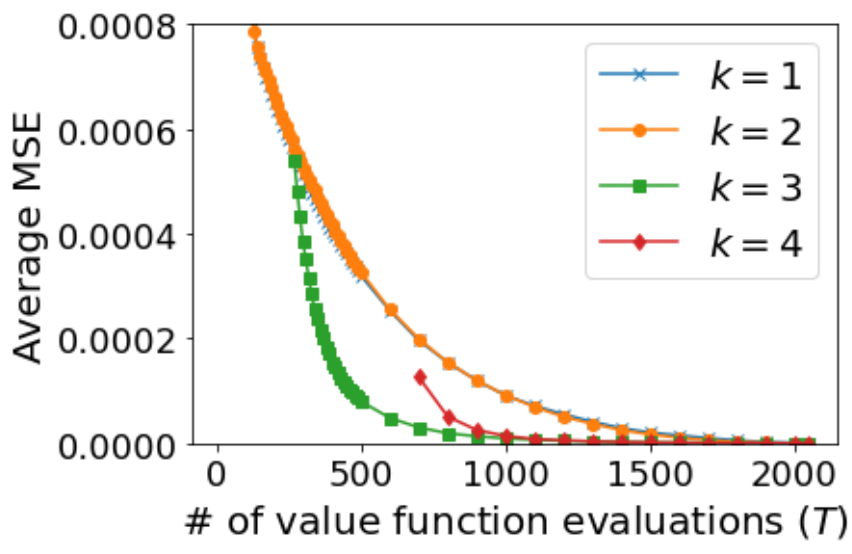}
    (a) Titanic ($n=11$)
    \label{fig:titanic_kadd}
\end{minipage}
\begin{minipage}[c]{0.28\textwidth}
    \centering
    \includegraphics[width=0.99\textwidth]{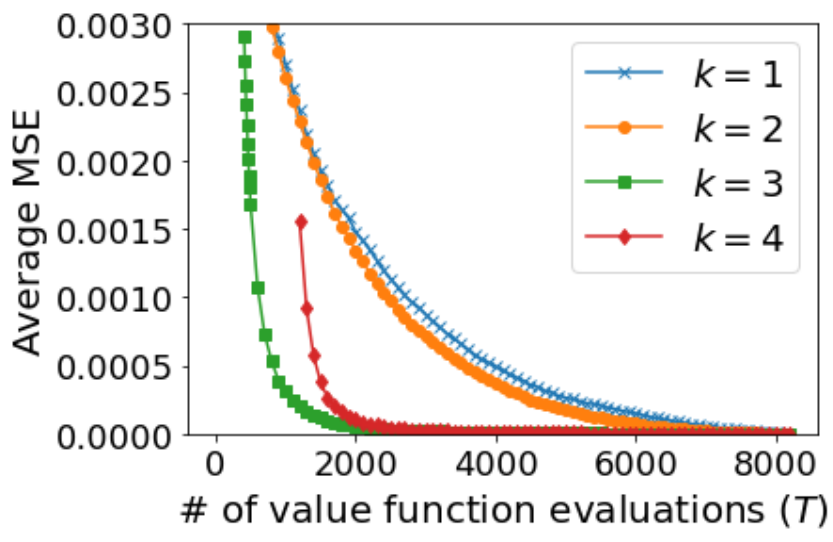}
    (b) Wine ($n=13$)
    \label{fig:wine_kadd}
\end{minipage}
\begin{minipage}[c]{0.28\textwidth}
    \centering
    \includegraphics[width=0.99\textwidth]{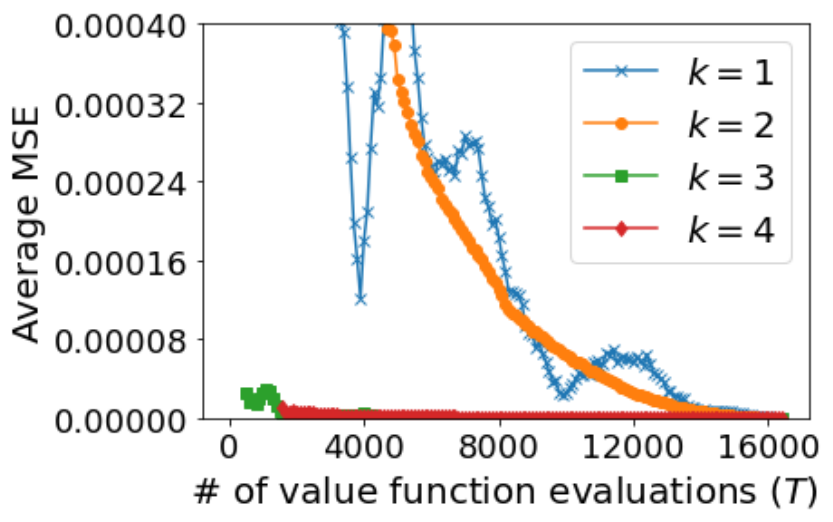}
    (c) Adult ($n=14$)
    \label{fig:adult_local_kadd}
\end{minipage}
\begin{minipage}[c]{0.28\textwidth}
    \centering
    \includegraphics[width=0.99\textwidth]{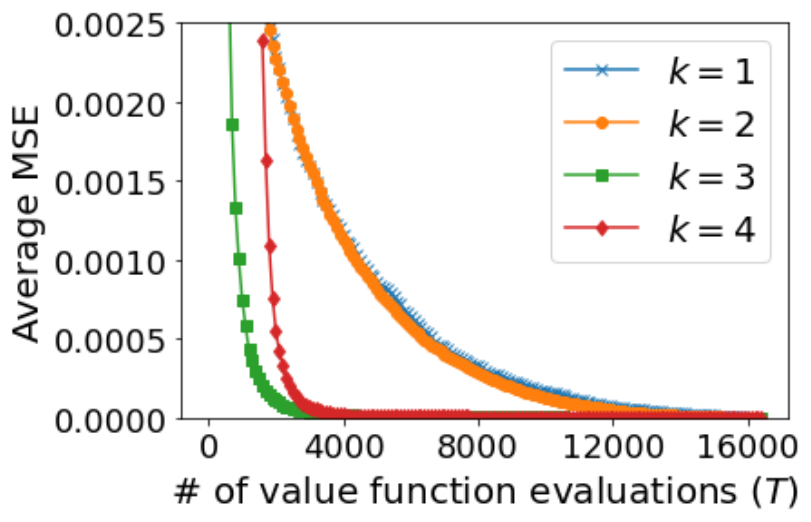}
    (d) ImageNet ($n=14$)
    \label{fig:image_cat_kadd}
\end{minipage}
\begin{minipage}[c]{0.28\textwidth}
    \centering
    \includegraphics[width=0.99\textwidth]{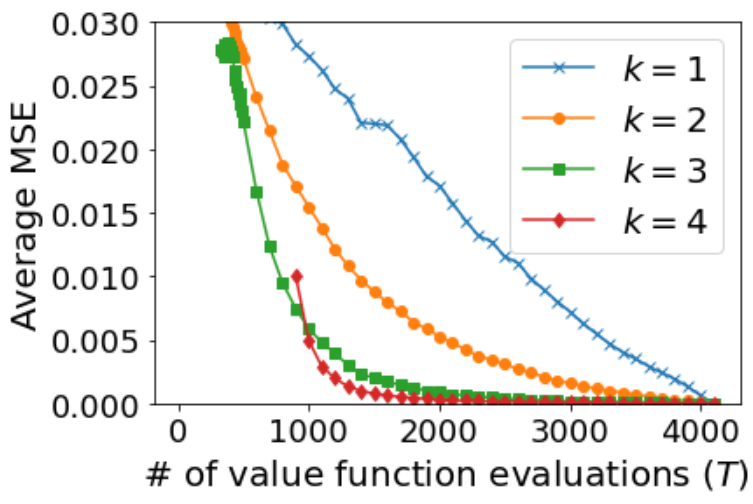}
    (e) Big Five ($n=12$)
    \label{fig:bigfive_kadd}
\end{minipage}
\begin{minipage}[c]{0.28\textwidth}
    \centering
    \includegraphics[width=0.99\textwidth]{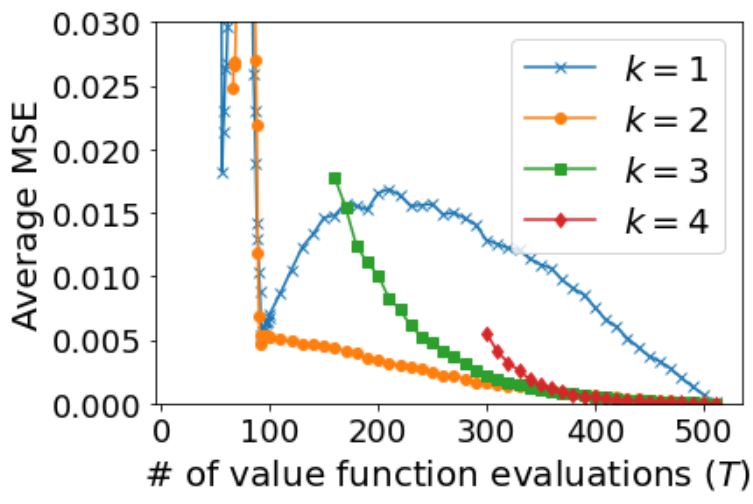}
    (f) Breast Cancer ($n=9$)
    \label{fig:breastcancer_kadd}
\end{minipage}
\caption{
MSE of \emph{SVA}$k_{\text{ADD}}$, averaged over 50 runs except for (c) and (d) with 10 runs, in dependence of available budget $T$ for different additivity degrees $k$.
Datasets stem from various explanation types: global (a)-(b), local (c)-(d), and unsupervised (e)-(f) with differing player numbers $n$.
}
\label{fig:proposal_results}
\end{figure*}

\section{Empirical Evaluation}
\label{sec:exper}

In order to assess the approximation performance of \emph{SVA}$k_{\text{ADD}}$, we conduct experiments with cooperative games stemming from various explanation types.
Although our method is not limited to a certain domain, we find approximating feature scores best to exemplarily illustrate its effectiveness on several datasets as a sanity check.
Our evaluation is mainly two-fold.
Not only are we interested in the comparison of \emph{SVA}$k_{\text{ADD}}$ against current state-of-the-art model-agnostic methods in \cref{subsec:degree}, but we also seek to investigate how the choice of the assumed degree of additivity $k$ affects the approximation quality (see \cref{subsec:performance}).
In the sequel of \cref{subsec:datasets}, we describe the utilized datasets and resulting cooperative games.

For each considered combination of dataset, approximation algorithm, and budget $T$, the obtained estimates $\hat\phi$ are compared with $\phi$ which we calculate exhaustively in advance.
We measure approximation quality of the estimates by the MSE.
The error is measured depending on $T$ as we intentionally refrain from a runtime comparison for multiple reasons:
(i) the observed runtimes may differ depending on the actual implementation,
(ii) evaluating the worth of a coalition poses the bottleneck in explanation tasks, rendering the difference in performed arithmetic operations negligible for more complex models and datasets,
(iii) instead of runtime, monetary units might be paid for each access to a remotely provided model offered by a third-party.

\subsection{Datasets}
\label{subsec:datasets}

We distinguish between three feature explanation tasks: global importance, local attribution, and unsupervised importance being described further in \cref{app:cooperative_games}.

Within global importance \citep{Covert.2020} the features' contributions to a model's generalization performance are quantified.
This is done by means of accuracy for classification and the MSE for regression~on~a test set.
For each evaluated coalition a random forest is trained on a training set.
We employ the \emph{Titanic} (classification, 11 features) and \emph{Wine} dataset (classification, 13  \mbox{features}).

On the contrary, local feature attribution \citep{Lundberg.2017} measures each feature's impact on the prediction of a fixed model for a given datapoint.
While the predicted value can directly be used as the worth of a feature coalition for regression, the predicted class probability is required instead of a label for classification. Rendering a feature outside of an evaluated coalition absent is performed by means of imputation that blurs the features contained information.
The experiments are conducted on the \emph{Adult} (classification, 14 features) and \emph{ImageNet} (classification, 14 features) dataset.

In the absence of labels, unsupervised feature importance \citep{Balestra.2022} seeks to find scores without a model's predictions.
This is achieved by employing the total correlation of a feature subset as its worth, since the datapoints can be seen as realizations of the joint feature value distribution.
For this task, we consider the \emph{Big Five} (12 features) and \emph{Breast Cancer} (9 features) datasets.

\begin{figure*}[t]
\centering
\begin{minipage}[c]{0.28\textwidth}
    \centering
    \includegraphics[width=0.99\textwidth]{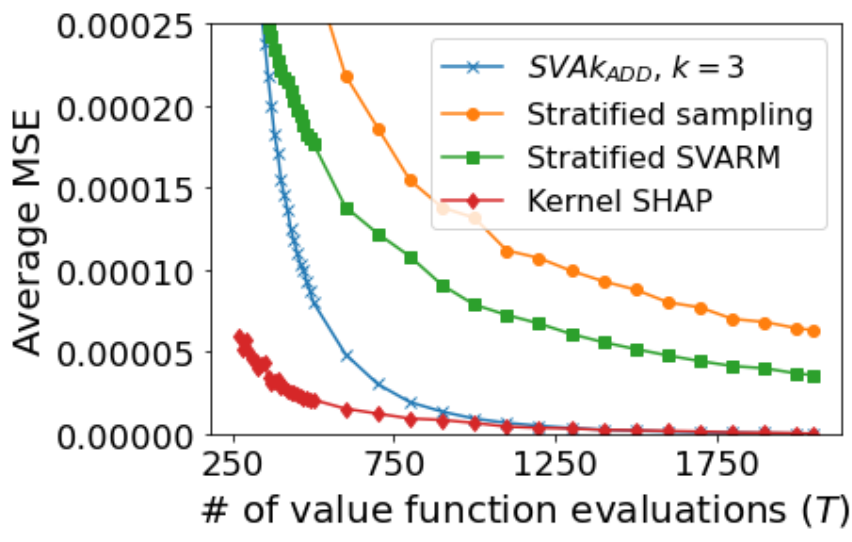}
    (a) Titanic dataset ($n=11$)
    \label{fig:titanic_kadd_comp}
\end{minipage}
\begin{minipage}[c]{0.28\textwidth}
    \centering
    \includegraphics[width=0.99\textwidth]{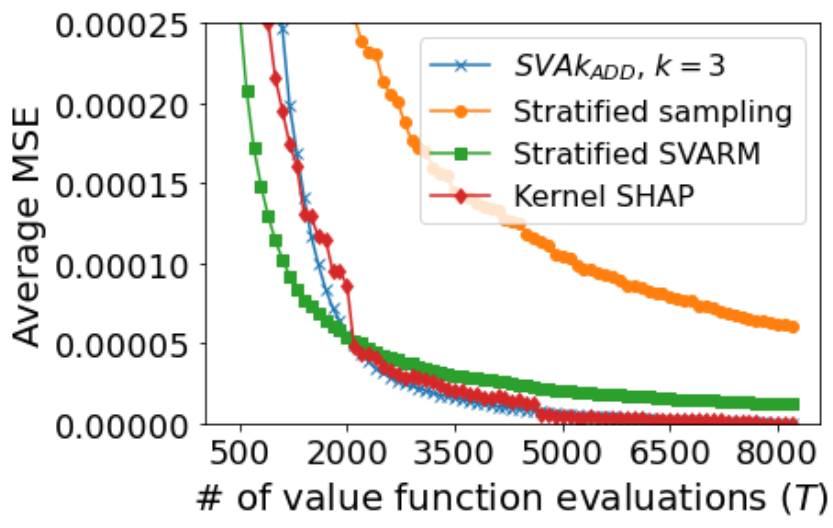}
    (b) Wine ($n=13$)
    \label{fig:wine_kadd_comp}
\end{minipage}
\begin{minipage}[c]{0.28\textwidth}
    \centering
    \includegraphics[width=0.99\textwidth]{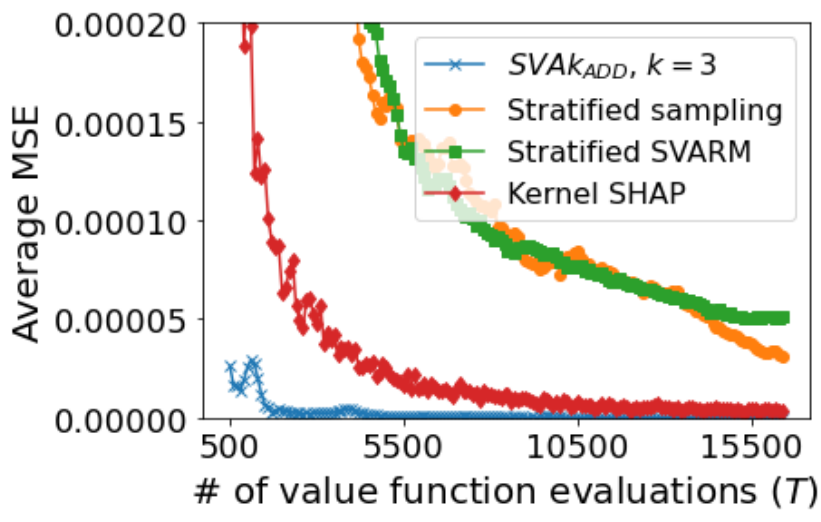}
    (c) Adult ($n=14$)
    \label{fig:adult_local_kadd_comp}
\end{minipage}
\begin{minipage}[c]{0.28\textwidth}
    \centering
    \includegraphics[width=0.99\textwidth]{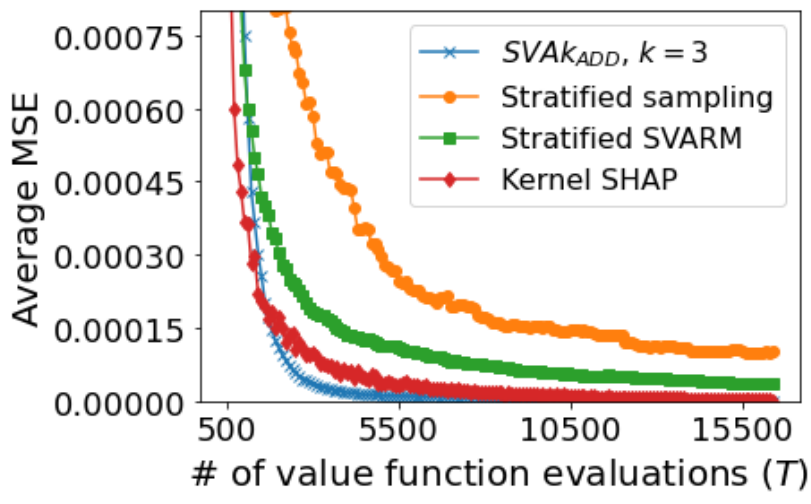}
    (d) ImageNet ($n=14$)
    \label{fig:image_cat_kadd_comp}
\end{minipage}
\begin{minipage}[c]{0.28\textwidth}
    \centering
    \includegraphics[width=0.99\textwidth]{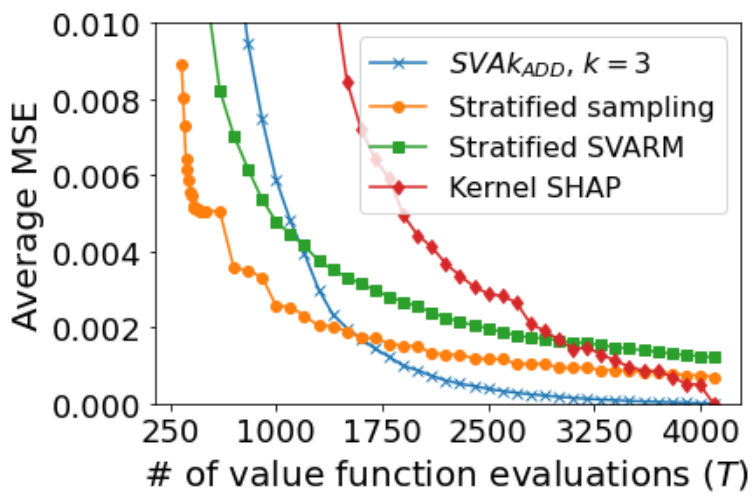}
    (e) Big Five ($n=12$)
    \label{fig:bigfive_kadd_comp}
\end{minipage}
\begin{minipage}[c]{0.28\textwidth}
    \centering
    \includegraphics[width=0.99\textwidth]{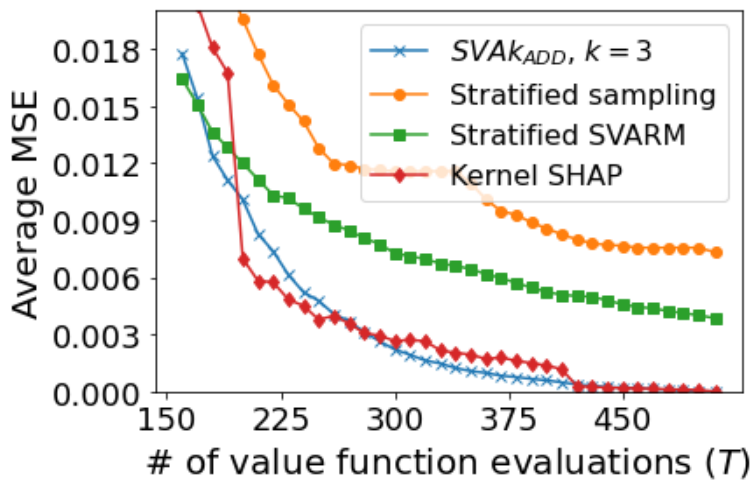}
    (f) Breast Cancer ($n=9$)
    \label{fig:breastcancer_kadd_comp}
\end{minipage}
\caption{
MSE of \emph{SVA}$k_{\text{ADD}}$ and competing methods, averaged over 50 runs except for (c) and (d) with 10 runs, in dependence of available budget $T$.
Datasets stem from various explanation types: global (a)-(b), local (c)-(d), and unsupervised (e)-(f) with differing player numbers $n$.
}
\label{fig:results_comp}
\end{figure*}

\subsection{Impact of the Additivity Degree $k$}
\label{subsec:degree}

In order to provide an understanding of the underlying trade-off between fast convergence (low $k$) and expressiveness (high $k$) of the surrogate game and how the crucial choice of $k$ affects the approximation quality, we evaluate \emph{SVA}$k_{\text{ADD}}$ for different $k \in \{1,2,3,4\}$ in Figure~\ref{fig:proposal_results}.
See Appendix~\ref{app:other_datasets} for more results with the \emph{Diabetes}, \emph{ImageNet}, and \emph{Breast cancer} dataset.

The curves for higher $k$ begin at points of higher budget because the greater $k$, the more coalition values are required to identify a unique $k$-additive value function that fits the observations.
We explain the behavior for low $k$, specifically $k=1$ and $k=2$, by the model's inability to achieve a good fit due to missing flexibility.
As a result, the convergence to the exact Shapley values is slow.
These findings imply that interactions up to order 2 are not sufficient to model how features jointly impact performance (global task) or prediction outcome (local task).
On the other hand, both the 3-additive and 4-additive model converge significantly faster for most datasets and outperform the parameterization with $k=1$ or $k=2$ after a few samples.
The choice of $k=3$ appears preferable as it results in quicker decreasing error curves.

\paragraph{Choice of $k$.}
The problem of choosing the optimal $k$ is intriguing and could potentially lead to further fruitful research.
It is comparable to related problems in other domains, for example finding the best $k$ in $k$-means clustering, and methods used there may also apply to our case.
A well-known example is the elbow heuristic: One fits multiple $k$-additive surrogate games, with increasing value for $k$, and monitors the resulting approximation error.
To our advantage, these multiple games can be fitted with one single stream of sampled coalitions, and hence we do not reduce the effectively available budget $T$.
The fit must improve monotonically for increasing $k$ (just like the objective function in k-means can only decrease with increasing $k$), because a larger $k$ only adds additional free variables in the optimization problem.
However, one typically observes bigger gains in the beginning (where $k$ is small), which at some point start to become much smaller.
Beyond this point (the ``elbow''), the additional gains by increasing $k$ are only marginal, so the elbow determines a good value for~$k$. 

\subsection{Comparison with Existing Methods}
\label{subsec:performance}

In our second experiment, we compare \emph{SVA}$k_{\text{ADD}}$ with other existing approximation methods.
For instance, we consider \emph{Stratified sampling} \cite{Maleki.2013}, \emph{Stratified SVARM} \cite{Kolpaczki.2024a} and \emph{KernelSHAP} \citep{Lundberg.2017}.
For the purpose of comparison, we adopt the $3$-additive model to represent \emph{SVA}$k_{\text{ADD}}$ since it displays the most satisfying compromise between approximation quality and minimum required evaluations as argued in \cref{subsec:degree}.
Figure~\ref{fig:results_comp} presents the obtained results for all methods.
See Appendix~\ref{app:futher_comparison} for more results with the \emph{Diabetes}, \emph{IMDB} and \emph{FIFA 21} dataset, including \emph{Permutation sampling} \cite{Castro.2009} and the $2$-additive model.

First to mention is that \emph{SVA}$k_{\text{ADD}}$ competes consistently with \emph{Stratified SVARM} and \emph{KernelSHAP} for the best approximation performance across all datasets.
Although for a very low number of function evaluations \emph{SVA}$k_{\text{ADD}}$ achieves an error greater than some other approaches,
at some point during the approximation process it converges faster to the exact Shapley values and leaves it competitors with a considerable margin behind, especially for local feature attribution on the \emph{Adult} and \emph{ImageNet} dataset.
In comparison to \emph{KernelSHAP} our method \emph{SVA}$k_{\text{ADD}}$ converges faster to the exact Shaley values for the \emph{Adult}, \emph{Big Five}, \emph{ImageNet}, and \emph{Wine} dataset, whereas for the \emph{Titanic} and \emph{Breast Cancer} dataset, \emph{KernelSHAP} achieves a better performance to which \emph{SVA}$k_{\text{ADD}}$ catches up with sufficient budget.

\section{Conclusion}
\label{sec:conclusion}

We proposed with \emph{SVA}$k_{\text{ADD}}$ a new algorithm to approximate Shapley values that fits a structured surrogate game instead of providing mean estimates via Monte Carlo sampling.
Despite restricting the surrogate game to be $k$-additive, our developed method is model-agnostic.
It is also applicable to any cooperative game without posing further assumptions since its underlying optimization problem provably yields the Shapley value.
We investigated empirically the trade-off that the choice of the parameter $k$ poses.
Further, \emph{SVA}$k_{\text{ADD}}$ exhibits competitive results with other existing approaches depending on the considered explanation type, dataset, and available budget, allowing us to conclude the non-existence of a dominating approximation method.

\paragraph{Limitations and future work.}

While the surrogate game's flexibility increases with higher $k$-additivity, it also requires more observations to obtain a unique solution of the optimization problem, eventually posing a practical limit on $k$.
The choice of $k$ is seemingly non-trivial and employing the elbow heuristic could provide a potential solution.
As the fit must improve monotonically for increasing $k$, one could fit multiple $k$-additive surrogate games and select the $k$ at the elbow where the gains in fit start to become significantly smaller. 
We expect future investigations of differently structured surrogate games to yield likewise fruitful results and contribute to the advancement of this class of approximation algorithms. 
Besides the estimated Shapley values, our proposal could also provide interaction effects when $k \geq 2$.
Although we did not address these parameters, future works can extract the estimated interactions to investigate redundant or complementary features.
This could be of interest in practical applications where interaction between features are relevant as for example in disease detection.

\section*{Acknowledgments}

This work was supported by the São Paulo Research Foundation (FAPESP, grant number 2025/00700-0)
and the German Research Foundation (Deutsche Forschungsgemeinschaft, DFG, TRR 318/1 2021 – 438445824).

\bibliography{references}

\appendix
\onecolumn
\newpage
\section{Theoretical Analysis} \label{app:analysis}

In the following we prove \cref{the:solution} 
by solving the $k$-additive optimization problem with weights $w_A = \binom{n-2}{|A|-1}^{-1}$ analytically and showing that the solution contains the Shapley value.
We introduce some simplifying notation:
\begin{itemize}
    \item $\tilde{\mathcal{P}} := \mathcal{P}(N) \setminus \{\emptyset,N\}$
    \item $I_0 := I^k(\emptyset)$, \hspace{0.1cm} $I_i := I^k(\{i\})$ for all $i \in N$, \hspace{0.1cm} $I_{i,j} := I^k(\{i,j\})$ for all $\{i,j\} \subseteq N$, \hspace{0.1cm} $I_{i,j,\ell} := I^k(\{i,j,\ell\})$
    \item Weight $w_a$ for any $A$ with $|A|=a$
    \item $\gamma_{A,B} := \gamma_{\left| A \cap B \right|}^{\left|B \right|}$ for all $A,B \subseteq N$
    \item $\beta_{i,A} = \begin{cases}
        1 & \text{if } i \notin A \\
        -1 & \text{if } i \in A
    \end{cases}$ \quad for all $i \in N$ and $A \in \mathcal{P}(N) \setminus \{\emptyset,N\}$
    \item $\beta_{i,j,A} = \begin{cases}
        2 & \text{if } |\{i,j\} \cap A| = 1 \\
        - 1 & \text{otherwise} \\
    \end{cases}$ \quad for all $\{i,j\} \subseteq N$ and $A \in \mathcal{P}(N) \setminus \{\emptyset,N\}$
    \item $\beta_{i,j,\ell,A} = \begin{cases}
        - 1 & \text{if } |\{i,j,\ell\} \cap A| = 1 \\
        1 & \text{if } |\{i,j,\ell\} \cap A| = 2 \\
        0 & \text{otherwise}
    \end{cases}$ \quad for all $\{i,j,\ell\} \subseteq N $ and $A \in \mathcal{P}(N) \setminus \{\emptyset,N\}$
\end{itemize}
Our proof is preceded by an observation that we shall utilize later:
\begin{lemma} \label{lem:sums}
    For any set of players $N$, player $i \in N$ and for the cases $\ell=2$ and $\ell=3$ the following equality holds:
    \begin{equation*}
        \sum\limits_{A \in \tilde{\mathcal{P}}} w_A \beta_{i,A} \sum\limits_{\substack{B \subseteq N \\ |B| = \ell}} \gamma_{A,B} I_B = \frac{1}{n} \sum\limits_{j \in N} \sum\limits_{A \in \tilde{\mathcal{P}}} w_A \beta_{j,A} \sum\limits_{\substack{B \subseteq N \\ |B| = \ell}} \gamma_{A,B} I_B \, .
    \end{equation*}
\end{lemma}
\noindent
\textbf{Proof:} \\
We show the statement for both cases separately and start with $\ell=2$.
For interactions $I_{j_1,j_2}$ that contain $i$ we derive after inserting the weights $w_a = \binom{n-2}{a-1}^{-1}$:
\begin{equation} \label{eq:pairs_i_included}
	\begin{array}{rll}
		& \ \sum\limits_{A \in \tilde{\mathcal{P}}} w_A \beta_{i,A} \sum\limits_{\substack{\{j_1,j_2\} \subseteq N \\ i \in \{j_1,j_2\}}} \beta_{j_1,j_2,A} I_{j_1,j_2} \\
        = & \ \sum\limits_{j_1 \in N \setminus \{i\}} I_{i,j_1} \sum\limits_{A \in \tilde{\mathcal{P}}} \beta_{i,A} \beta_{j_1,i,A} w_A \\
        = & \ \sum\limits_{j_1 \in N \setminus \{i\}} I_{i,j_1} \left( \sum\limits_{\substack{A \in \tilde{\mathcal{P}} \\ i,j_1 \in A}} w_A - \sum\limits_{\substack{A \in \tilde{\mathcal{P}} \\ i,j_1 \notin A}} w_A -2 \sum\limits_{\substack{A \in \tilde{\mathcal{P}} \\ i \in A, j_1 \notin A}} w_A + 2 \sum\limits_{\substack{A \in \tilde{\mathcal{P}} \\ i \notin A, j_1 \in A}} w_A \right) \\
        = & \ \sum\limits_{j_1 \in N \setminus \{i\}} I_{i,j_1} \left( \sum\limits_{a=2}^{n-1} \binom{n-2}{a-2} w_a - \sum\limits_{a=1}^{n-2} \binom{n-2}{a} w_a -2 \sum\limits_{a=1}^{n-1} \binom{n-2}{a-1} w_a + 2 \sum\limits_{a=1}^{n-1} \binom{n-2}{a-1} w_a \right) \\
        = & \ \sum\limits_{j_1 \in N \setminus \{i\}} I_{i,j_1} \sum\limits_{a=1}^{n-1} \left( \binom{n-2}{a-2} - \binom{n-2}{a} \right) w_a \\
        = & \ \sum\limits_{j_1 \in N, j_1 \neq i} I_{j_1,i} \sum\limits_{a=1}^{n-1} \frac{a-1}{n-a} - \frac{n-a-1}{a} \\
        = & \ 0
	\end{array}
\end{equation}
And for all other interactions $I_{j_1,j_2}$ not containing $i$ we derive:
\begin{equation} \label{eq:pairs_i_excluded}
	\begin{array}{rll}
		& \ \sum\limits_{A \in \tilde{\mathcal{P}}} w_A \beta_{i,A} \sum\limits_{\{j_1,j_2\} \subseteq N \setminus \{i\}} \beta_{j_1,j_2,A} I_{j_1,j_2} \\
        = & \ \sum\limits_{\{j_1,j_2\} \subseteq N \setminus \{i\}} I_{j_1,j_2} \Bigg( \sum\limits_{\substack{A \in \tilde{\mathcal{P}} \\ i,j_1,j_2 \in A}} w_A + \sum\limits_{\substack{A \in \tilde{\mathcal{P}} \\ i \in A, j_1,j_2 \notin A}} w_A - 2\sum\limits_{\substack{A \in \tilde{\mathcal{P}} \\ i,j_1 \in A, j_2 \notin A}} w_A - 2\sum\limits_{\substack{A \in \tilde{\mathcal{P}} \\ i,j_2 \in A, j_1 \notin A}} w_A \\
        & \ - \sum\limits_{\substack{A \in \tilde{\mathcal{P}} \\ i \notin A, j_1,j_2 \in A}} w_A - \sum\limits_{\substack{A \in \tilde{\mathcal{P}} \\ i,j_1,j_2 \notin A}} w_A + 2\sum\limits_{\substack{A \in \tilde{\mathcal{P}} \\ j_1 \in A, i,j_2 \notin A}} w_A + 2\sum\limits_{\substack{A \in \tilde{\mathcal{P}} \\ j_2 \in A, i,j_1 \notin A}} w_A \Bigg) \\
        = & \ \sum\limits_{\{j_1,j_2\} \subseteq N \setminus \{i\}} I_{j_1,j_2} \Bigg( \sum\limits_{a=3}^{n-1} \binom{n-3}{a-3} w_a + \sum\limits_{a=1}^{n-2} \binom{n-3}{a-1} w_a - 2\sum\limits_{a=2}^{n-1} \binom{n-3}{a-2} w_a - 2\sum\limits_{a=2}^{n-1} \binom{n-3}{a-2} w_a \\
        & \ - \sum\limits_{a=2}^{n-1} \binom{n-3}{a-2} w_a - \sum\limits_{a=1}^{n-3} \binom{n-3}{a} w_a + 2\sum\limits_{a=1}^{n-2} \binom{n-3}{a-1} w_a + 2\sum\limits_{a=1}^{n-2} \binom{n-3}{a-1} w_a \Bigg) \\
        = & \ \sum\limits_{\{j_1,j_2\} \subseteq N \setminus \{i\}} I_{j_1,j_2} \sum\limits_{a=1}^{n-1} \left( \binom{n-3}{a-3} -5 \binom{n-3}{a-2} + 5\binom{n-3}{a-1} - \binom{n-3}{a} \right) w_a \\
        = & \ \sum\limits_{\{j_1,j_2\} \subseteq N \setminus \{i\}} I_{j_1,j_2} \sum\limits_{a=1}^{n-1} \frac{(a-1)(a-2)}{(n-2)(n-a)} - 5\frac{a-1}{n-2} + 5\frac{n-a-1}{n-2} - \frac{(n-a-1)(n-a-2)}{(n-2)a}  \\
        = & \ 0
	\end{array}
\end{equation}
Adding \cref{eq:pairs_i_included} and (\ref{eq:pairs_i_excluded}) yields:
\begin{equation*}
	\begin{array}{rl}
        & \ \sum\limits_{A \in \tilde{\mathcal{P}}} w_A \beta_{i,A} \sum\limits_{\substack{B \subseteq N \\ |B| = 2}} \gamma_{A,B} I_B \\
        = & \ - \frac{1}{6} \sum\limits_{A \in \tilde{\mathcal{P}}} w_A \beta_{i,A} \sum\limits_{\substack{\{j_1,j_2\} \subseteq N}} \beta_{j_1,j_2,A} I_{j_1,j_2} \\
        = & \ - \frac{1}{6} \left( \sum\limits_{A \in \tilde{\mathcal{P}}} w_A \beta_{i,A} \sum\limits_{\substack{\{j_1,j_2\} \subseteq N \\ i \in \{j_1,j_2\}}} \beta_{j_1,j_2,A} I_{j_1,j_2} + \sum\limits_{A \in \tilde{\mathcal{P}}} w_A \beta_{i,A} \sum\limits_{\{j_1,j_2\} \subseteq N \setminus \{i\}} \beta_{j_1,j_2,A} I_{j_1,j_2} \right) \\
        = & \ 0
	\end{array}
\end{equation*}
Consequently, we also have
\begin{equation*}
	\begin{array}{rll}
        \frac{1}{n} \sum\limits_{j \in N} \sum\limits_{A \in \tilde{\mathcal{P}}} w_A \beta_{j,A} \sum\limits_{\substack{B \subseteq N \\ |B| = 2}} \gamma_{A,B} I_B = 0 \, .
	\end{array}
\end{equation*}
Continuing with $\ell=3$, for interactions $I_{j_1,j_2,j_3}$ that contain $i$ we derive after inserting the weights $w_a = \binom{n-2}{a-1}^{-1}$:
\begin{equation} \label{eq:triples_i_included}
	\begin{array}{rll}
		& \ \sum\limits_{A \in \tilde{\mathcal{P}}} w_A \beta_{i,A} \sum\limits_{\substack{\{j_1,j_2,j_3\} \subseteq N \\ i \in \{j_1,j_2,j_3\}}} \beta_{j_1,j_2,j_3,A} I_{j_1,j_2,j_3} \\
        = & \ \sum\limits_{\{j_1,j_2\} \subseteq N \setminus \{i\}} I_{i,j_1,j_2} \sum\limits_{A \in \tilde{P}} \beta_{i,A} \beta_{i,j_1,j_2} w_A \\
        = & \ \sum\limits_{\{j_1,j_2\} \subseteq N \setminus \{i\}} I_{i,j_1,j_2} \left( \sum\limits_{\substack{A \in \tilde{\mathcal{P}}, i \in A \\ |\{j_1,j_2\} \cap A|=0}} w_A - \sum\limits_{\substack{A \in \tilde{\mathcal{P}}, i \in A \\ |\{j_1,j_2\} \cap A|=1}} w_A - \sum\limits_{\substack{A \in \tilde{\mathcal{P}}, i \notin A \\ |\{j_1,j_2\} \cap A|=1}} w_A + \sum\limits_{\substack{A \in \tilde{\mathcal{P}}, i \notin A \\ |\{j_1,j_2\} \cap A|=2}} w_A \right) \\
        = & \ \sum\limits_{\{j_1,j_2\} \subseteq N \setminus \{i\}} I_{i,j_1,j_2} \left( \sum\limits_{a=1}^{n-2} \binom{n-3}{a-1} w_a - 2\sum\limits_{a=2}^{n-1} \binom{n-3}{a-2} w_a - 2\sum\limits_{a=1}^{n-2} \binom{n-3}{a-1} w_a + \sum\limits_{a=2}^{n-1} \binom{n-3}{a-2} w_a \right) \\
        = & \ - \sum\limits_{\{j_1,j_2\} \subseteq N \setminus \{i\}} I_{i,j_1,j_2} \sum\limits_{a=1}^{n-1} \binom{n-2}{a-1} w_a \\
        = & \ - (n-1) \sum\limits_{\{j_1,j_2\} \subseteq N \setminus \{i\}} I_{i,j_1,j_2}
	\end{array}
\end{equation}
And for all other interactions $I_{j_1,j_2,j_3}$ not containing $i$ we derive:
\begin{equation} \label{eq:triples_i_excluded}
	\begin{array}{rll}
		& \ \sum\limits_{A \in \tilde{\mathcal{P}}} w_A \beta_{i,A} \sum\limits_{\{j_1,j_2,j_3\} \subseteq N \setminus \{i\}} \beta_{j_1,j_2,j_3,A} I_{j_1,j_2,j_3} \\
        = & \ \sum\limits_{\{j_1,j_2,j_3\} \subseteq N \setminus \{i\}} I_{j_1,j_2,j_3} \sum\limits_{A \in \tilde{\mathcal{P}}} \beta_{i,A} \beta_{j_1,j_2,j_3,A} w_A \\
        = & \ \sum\limits_{\{j_1,j_2,j_3\} \subseteq N \setminus \{i\}} I_{j_1,j_2,j_3} \left( \sum\limits_{\substack{A \in \tilde{\mathcal{P}}, i \in A \\ |\{j_1,j_2,j_3\} \cap A| = 1}} w_A - \sum\limits_{\substack{A \in \tilde{\mathcal{P}}, i \in A \\ |\{j_1,j_2,j_3\} \cap A| = 2}} w_A - \sum\limits_{\substack{A \in \tilde{\mathcal{P}}, i \notin A \\ |\{j_1,j_2,j_3\} \cap A| = 1}} w_A + \sum\limits_{\substack{A \in \tilde{\mathcal{P}}, i \notin A \\ |\{j_1,j_2,j_3\} \cap A| = 2}} w_A \right) \\
        = & \ 3 \sum\limits_{\{j_1,j_2,j_3\} \subseteq N \setminus \{i\}} I_{j_1,j_2,j_3} \left( \sum\limits_{a=2}^{n-2} \binom{n-4}{a-2} w_a - \sum\limits_{a=3}^{n-1} \binom{n-4}{a-3} w_a - \sum\limits_{a=1}^{n-3} \binom{n-4}{a-1} w_a + \sum\limits_{a=2}^{n-2} \binom{n-4}{a-2} w_a \right) \\
        = & \ 3 \sum\limits_{\{j_1,j_2,j_3\} \subseteq N \setminus \{i\}} I_{j_1,j_2,j_3} \sum\limits_{a=1}^{n-1} \left( 4\binom{n-4}{a-2} - \binom{n-2}{a-1} \right) w_a \\
        = & \ 3 \sum\limits_{\{j_1,j_2,j_3\} \subseteq N \setminus \{i\}} I_{j_1,j_2,j_3} \sum\limits_{a=1}^{n-1} 4 \frac{(a-1)(n-a-1)}{(n-2)(n-3)} - 1 \\
        = & \ -(n-1) \sum\limits_{\{j_1,j_2,j_3\} \subseteq N \setminus \{i\}} I_{j_1,j_2,j_3}
	\end{array}
\end{equation}
Adding \cref{eq:triples_i_included} and (\ref{eq:triples_i_excluded}) yields:
\begin{equation*}
	\begin{array}{rl}
        & \ \sum\limits_{A \in \tilde{\mathcal{P}}} w_A \beta_{i,A} \sum\limits_{\substack{B \subseteq N \\ |B| = 3}} \gamma_{A,B} I_B \\
        = & \ - \frac{1}{6} \sum\limits_{A \in \tilde{\mathcal{P}}} w_A \beta_{i,A} \sum\limits_{\substack{\{j_1,j_2,j_3\} \subseteq N}} \beta_{j_1,j_2,j_3,A} I_{j_1,j_2,j_3} \\
        = & \ - \frac{1}{6} \left( \sum\limits_{A \in \tilde{\mathcal{P}}} w_A \beta_{i,A} \sum\limits_{\substack{\{j_1,j_2,j_3\} \subseteq N \\ i \in \{j_1,j_2,j_3\}}} \beta_{j_1,j_2,j_3,A} I_{j_1,j_2,j_3} + \sum\limits_{A \in \tilde{\mathcal{P}}} w_A \beta_{i,A} \sum\limits_{\{j_1,j_2,j_3\} \subseteq N \setminus \{i\}} \beta_{j_1,j_2,j_3,A} I_{j_1,j_2,j_3} \right) \\
        = & \ \frac{n-1}{6} \left( \sum\limits_{\{j_1,j_2\} \subseteq N \setminus \{i\}} I_{i,j_1,j_2} + \sum\limits_{\{j_1,j_2,j_3\} \subseteq N \setminus \{i\}} I_{j_1,j_2,j_3} \right) \\
        = & \ \frac{n-1}{6} \sum\limits_{\{j_1,j_2,j_3\} \subseteq N} I_{j_1,j_2,j_3}
	\end{array}
\end{equation*}
Obviously, summing up the last term over all $j \in N$ and dividing it by $n$ will not change it, which concludes the proof. \qed

\noindent \\
\textbf{Proof of \cref{the:solution}:} \\
The constraint to guarantee the efficiency axiom can be simplified, leading to the following optimization problem:
\begin{equation*}
	\begin{array}{rl}
		\displaystyle\min_I & \sum\limits_{A \in \tilde{\mathcal{P}}} w_A \left( \nu(A) - \sum\limits_{B \subseteq N, |B| \leq k} \gamma_{A,B} I_B \right)^2 \\
		\text{s.t.} & \nu(N) - \nu(\emptyset) \ = \sum\limits_{i \in N} I_i
	\end{array}
\end{equation*}
We apply the Lagrange method.
The new objective to minimize is
\begin{equation*}
	\begin{array}{rl}
		\Lambda(I, \lambda) = \sum\limits_{A \in \tilde{\mathcal{P}}} w_A \left( \nu(A) - \sum\limits_{B \subseteq N, |B| \leq k} \gamma_{A,B} I_B \right)^2 + \lambda \left( \sum\limits_{i \in N} I_i - \nu(N) + \nu(\emptyset) \right) \, .
	\end{array}
\end{equation*}
The partial derivatives of $\Lambda$ must turn to zero for its solution.
Hence we obtain the following equations:
\begin{equation*}
	\begin{array}{rll}
        \frac{\partial \Lambda}{\partial I_0} = & - 2 \sum\limits_{A \in \tilde{\mathcal{P}}} w_A \gamma_{A,\emptyset} \left( \nu(A) - \sum\limits_{B \subseteq N, |B| \leq k} \gamma_{A,B} I_B \right) & \stackrel{!}{=} 0 \\
        \frac{\partial \Lambda}{\partial I_i} = & - 2 \sum\limits_{A \in \tilde{\mathcal{P}}} w_A \gamma_{A,\{i\}} \left( \nu(A) - \sum\limits_{B \subseteq N, |B| \leq k} \gamma_{A,B} I_B \right) + \lambda & \stackrel{!}{=} 0 \hspace{0.1cm} \text{ for all } i \in N \\
        \frac{\partial \Lambda}{\partial I_S} = & - 2 \sum\limits_{A \in \tilde{\mathcal{P}}} w_A \gamma_{A,S} \left( \nu(A) - \sum\limits_{B \subseteq N, |B| \leq k} \gamma_{A,B} I_B \right) & \stackrel{!}{=} 0 \hspace{0.1cm} \text{ for all } S \subseteq N \text{ with } |S| \in [2,k] \\
        \frac{\partial \Lambda}{\partial \lambda} = & \sum\limits_{i \in N} I_i - \nu(N) + \nu(\emptyset) & \stackrel{!}{=} 0
	\end{array}
\end{equation*}
From $\frac{\partial \Lambda}{\partial \lambda}$ we immediately extract
\begin{equation} \label{eq:efficiency_sum}
    \begin{array}{rll}
		\sum\limits_{i \in N} I_i = \nu(N) - \nu(\emptyset)
	\end{array}
\end{equation}
and thus also for any $i \in N$:
\begin{equation} \label{eq:efficiency_excluded}
    \begin{array}{rll}
        \sum\limits_{j \in N \setminus \{i\}} I_j = \nu(N) - \nu(\emptyset) - I_i \,.
    \end{array}
\end{equation}
The derivative $\frac{\partial \Lambda}{\partial I_i}$ can be rearranged for any $i \in N$ to obtain an expression of summands grouped by the order of their contained interactions:
\begin{equation} \label{eq:derivative}
	\begin{array}{rl}
		\frac{\partial \Lambda}{\partial I_i} & = \sum\limits_{A \in \tilde{\mathcal{P}}} w_A \beta_{i,A} \left( \nu(A) - \sum\limits_{B \subseteq N, |B| \leq k} \gamma_{A,B} I_B \right) + \lambda \\
        & = \sum\limits_{A \in \tilde{\mathcal{P}}} w_A \beta_{i,A} \nu(A) - \sum\limits_{A \in \tilde{\mathcal{P}}} w_A \beta_{i,A} I_0 + \frac{1}{2} \sum\limits_{A \in \tilde{\mathcal{P}}} w_A \beta_{i,A} \sum\limits_{j \in N}\beta_{j,A} I_j \\
        & \quad - \sum\limits_{A \in \tilde{\mathcal{P}}} w_A \beta_{i,A} \sum\limits_{\substack{B \subseteq N \\ 2 \leq |B| \leq k}} \gamma_{A,B} I_B + \lambda \\
	\end{array}
\end{equation}
In the following, we derive expressions for multiple terms that are contained in \cref{eq:derivative}.
First, we have for the sum containing the interaction of the empty set:
\begin{equation} \label{eq:Izero}
	\begin{array}{rll}
		\sum\limits_{A \in \tilde{\mathcal{P}}} w_A \beta_{i,A} I_0 & = I_0 \left( \sum\limits_{A \in \tilde{\mathcal{P}}, i \notin A} w_A  - \sum\limits_{A \in \tilde{\mathcal{P}}, i \in A} w_A \right) \\
        & = I_0 \left( \sum\limits_{a=1}^{n-1} \sum\limits_{\substack{A \in \tilde{\mathcal{P}}, i \notin A \\ |A|=a}} w_a  - \sum\limits_{a=1}^{n-1} \sum\limits_{\substack{A \in \tilde{\mathcal{P}}, i \in A \\ |A|=a}} w_a \right) \\
        & = I_0 \sum\limits_{a=1}^{n-1} \left( \binom{n-1}{a} - \binom{n-1}{a-1} \right) w_a \\
        & = I_0 \sum\limits_{a=1}^{n-1} \frac{n-1}{a} - \frac{n-1}{n-a} \\
        & = 0
	\end{array}
\end{equation}
Next, we solve the sum that contains interactions of singletons, requiring two steps, we begin with
\begin{equation} \label{eq:shapleys_same}
    \begin{array}{rll}
        \sum\limits_{A \in \tilde{\mathcal{P}}} w_A \beta_{i,A}^2 I_i = I_i \sum\limits_{a=1}^{n-1} \binom{n}{a} w_a \, .
    \end{array}
\end{equation}
In the second step we analyze and utilize \cref{eq:efficiency_excluded} to obtain:
\begin{equation} \label{eq:shapleys_different}
	\begin{array}{rll}
		& \ \sum\limits_{A \in \tilde{\mathcal{P}}} w_A \beta_{i,A} \sum\limits_{j \in N, j \neq i} \beta_{j,A} I_j \\
        = & \ \sum\limits_{j \in N, j \neq i} I_j \Bigg( \sum\limits_{\substack{A \in \tilde{\mathcal{P}} \\ i,j \in A}} w_A + \sum\limits_{\substack{A \in \tilde{\mathcal{P}} \\ i,j \notin A}} w_A - \sum\limits_{\substack{A \in \tilde{\mathcal{P}} \\ i \in A, j \notin A}} w_A - \sum\limits_{\substack{A \in \tilde{\mathcal{P}} \\ i \notin A, j \in A}} w_A \Bigg) \\
        = & \ \sum\limits_{j \in N, j \neq i} I_j \left( \sum\limits_{a=2}^{n-1} \binom{n-2}{a-2} w_a + \sum\limits_{a=1}^{n-2} \binom{n-2}{a} w_a - 2\sum\limits_{a=1}^{n-1} \binom{n-2}{a-1} w_a \right) \\
        = & \ \left( \nu(N) - \nu(\emptyset) - I_i \right) \sum\limits_{a=1}^{n-1} \left( \binom{n-2}{a-2} + \binom{n-2}{a} -2 \binom{n-2}{a-1} \right) w_a
	\end{array}
\end{equation}
We combine \cref{eq:shapleys_same} with \cref{eq:shapleys_different}, and apply the weights $w_a = \binom{n-2}{a-1}^{-1}$ and the identity $2H_{n-1} = \sum\nolimits_{a=1}^{n-1} \frac{n}{a(n-a)}$, where $H_{n-1} = \sum\nolimits_{a=1}^{n-1} \frac{1}{a}$ is the harmonic sum, to derive:
\begin{equation} \label{eq:shapleys_all}
	\begin{array}{rll}
		& \ \sum\limits_{A \in \tilde{\mathcal{P}}} w_A \beta_{i,A} \sum\limits_{j \in N} \beta_{j,A} I_j \\
        = & \ \sum\limits_{A \in \tilde{\mathcal{P}}} w_A \beta_{i,A} \left( \beta_{i,A} I_i + \sum\limits_{j \in N, j \neq i} \beta_{j,A} I_j \right) \\
        = & \ \sum\limits_{A \in \tilde{\mathcal{P}}} w_A \beta_{i,A}^2 I_i + \sum\limits_{A \in \tilde{\mathcal{P}}} w_A \beta_{i,A} \sum\limits_{j \in N, j \neq i} \beta_{j,A} I_j \\
        = & \ I_i \sum\limits_{a=1}^{n-1} \binom{n}{a} w_a + \left( \nu(N) - \nu(\emptyset) - I_i \right) \sum\limits_{a=1}^{n-1} \left( \binom{n-2}{a-2} + \binom{n-2}{a} -2 \binom{n-2}{a-1} \right) w_a \\
        = & \ I_i \sum\limits_{a=1}^{n-1} \left( \binom{n}{a} - \binom{n-2}{a-2} - \binom{n-2}{a} + 2\binom{n-2}{a-1} \right) w_a + \left( \nu(N) - \nu(\emptyset) \right) \sum\limits_{a=1}^{n-1} \left( \binom{n-2}{a-2} + \binom{n-2}{a} -2 \binom{n-2}{a-1} \right) w_a \\
        = & \ 4 I_i \sum\limits_{a=1}^{n-1} \binom{n-2}{a-1} w_a + \left( \nu(N) - \nu(\emptyset) \right) \sum\limits_{a=1}^{n-1} \left( \binom{n}{a} - 4\binom{n-2}{a-1} \right) w_a \\
        = & \ 4(n-1) I_i + \left( \nu(N) - \nu(\emptyset) \right) \sum\limits_{a=1}^{n-1} \left(\frac{n(n-1)}{a(n-a)} -4 \right) \\
        = & \ 4(n-1) I_i + 2(n-1)(H_{n-1} - 2) \left( \nu(N) - \nu(\emptyset) \right)
	\end{array}
\end{equation}
Summing \cref{eq:shapleys_all} up over all $i \in N$ yields under usage of \cref{eq:efficiency_sum}:
\begin{equation} \label{eq:shapleys_all_sum}
    \begin{array}{rll}
        \sum\limits_{i \in N} \sum\limits_{A \in \tilde{\mathcal{P}}} w_A \beta_{i,A} \sum\limits_{j \in N} \beta_{j,A} I_j = 2(n-1)(n(H_{n-1} - 2) + 2) \left( \nu(N) - \nu(\emptyset) \right) \, .
    \end{array}
\end{equation}
And as the final intermediate term we derive for the weighted coalition values summed up over all $i \in N$:
\begin{equation} \label{eq:weighted_values}
	\begin{array}{rll}
        \sum\limits_{i \in N} \sum\limits_{A \in \tilde{\mathcal{P}}} w_A \beta_{i,A} \nu(A) = \sum\limits_{A \in \tilde{\mathcal{P}}} w_A \nu(A) \left( n-2|A| \right) \, .
	\end{array}
\end{equation}
For any $i \in N$, after rearranging \cref{eq:derivative} and plugging in \cref{eq:Izero} and (\ref{eq:shapleys_all}) we have
\begin{equation} \label{eq:lambdas}
    \begin{array}{rl}
        - \lambda = & \ \sum\limits_{A \in \tilde{\mathcal{P}}} w_A \beta_{i,A} \nu(A) + 2(n-1) I_i + (n-1)(H_{n-1} - 2) \left( \nu(N) - \nu(\emptyset) \right) \\
        & \quad - \sum\limits_{A \in \tilde{\mathcal{P}}} w_A \beta_{i,A} \sum\limits_{\substack{B \subseteq N \\ 2 \leq |B| \leq k}} \gamma_{A,B} I_B \, .
    \end{array}
\end{equation}
We also obtain for $-\lambda$ by rearranging \cref{eq:derivative}, summing it up over all $i \in N$, dividing by $n$, and plugging in \cref{eq:Izero}, (\ref{eq:shapleys_all_sum}), and (\ref{eq:weighted_values}):
\begin{equation} \label{eq:lambdas_sum}
	\begin{array}{rll}
		- \lambda = & \ \frac{1}{n} \sum\limits_{A \in \tilde{\mathcal{P}}} w_A \nu(A) \left( n-2|A| \right) + \frac{n-1}{n} (n(H_{n-1} - 2) + 2) \left( \nu(N) - \nu(\emptyset) \right) \\
        & \quad - \frac{1}{n} \sum\limits_{i \in N} \sum\limits_{A \in \tilde{\mathcal{P}}} w_A \beta_{i,A} \sum\limits_{\substack{B \subseteq N \\ 2 \leq |B| \leq k}} \gamma_{A,B} I_B \, .
	\end{array}
\end{equation}
Finally, we conclude the proof by equating \cref{eq:lambdas} and (\ref{eq:lambdas_sum}).
We utilize \cref{lem:sums} to cancel out the sums that contain interactions of order two and three such that the theorem holds true for the cases of $k=2$ and $k=3$.
This step is does not apply for the special case of $k=1$ since the last sums in \cref{eq:lambdas} and (\ref{eq:lambdas_sum}) vanish.
We solve for $I_i$ and derive:
\begin{equation*}
	\begin{array}{rll}
        I_i = & \ \frac{1}{2n} (n(H_{n-1} - 2) + 2) \left( \nu(N) - \nu(\emptyset) \right) - \frac{H_{n-1} - 2}{2} \left( \nu(N) - \nu(\emptyset) \right) \\
        & \ + \frac{1}{2n(n-1)} \sum\limits_{A \in \tilde{\mathcal{P}}} w_A \nu(A) \left( n-2|A| \right) - \frac{1}{2(n-1)} \sum\limits_{A \in \tilde{\mathcal{P}}} w_A \beta_{i,A} \nu(A) \\
        & \ + \sum\limits_{A \in \tilde{\mathcal{P}}} w_A \beta_{i,A} \sum\limits_{\substack{B \subseteq N \\ 2 \leq |B| \leq k}} \gamma_{A,B} I_B - \frac{1}{n} \sum\limits_{j \in N} \sum\limits_{A \in \tilde{\mathcal{P}}} w_A \beta_{j,A} \sum\limits_{\substack{B \subseteq N \\ 2 \leq |B| \leq k}} \gamma_{A,B} I_B \\
        = & \ \frac{1}{n} \left( \nu(N) - \nu(\emptyset) \right) + \frac{1}{2n(n-1)} \sum\limits_{A \in \tilde{\mathcal{P}} i \in A} w_A \nu(A) \left( n-2|A| \right) + \frac{1}{2(n-1)} \sum\limits_{A \in \tilde{\mathcal{P}}, i \in A} w_A \nu(A) \\
        & \ + \frac{1}{2n(n-1)} \sum\limits_{A \in \tilde{\mathcal{P}}, i \notin A} w_A \nu(A) \left( n-2|A| \right) - \frac{1}{2(n-1)} \sum\limits_{A \in \tilde{\mathcal{P}}, i \notin A} w_A \nu(A) \\
        = & \ \frac{1}{n} \left( \nu(N) - \nu(\emptyset) \right) + \sum\limits_{A \in \tilde{\mathcal{P}}, i \in A} \frac{n-|A|}{n(n-1)} \cdot w_A \nu(A) - \sum\limits_{A \in \tilde{\mathcal{P}}, i \notin A} \frac{|A|}{n(n-1)} \cdot w_A \nu(A) \\
        = & \ \frac{1}{n} \left( \nu(N) - \nu(\emptyset) \right) + \sum\limits_{A \in \tilde{\mathcal{P}}, i \in A} \frac{1}{n \cdot \binom{n-1}{|A|-1}} \cdot \nu(A) - \sum\limits_{A \in \tilde{\mathcal{P}}, i \notin A} \frac{1}{n \cdot \binom{n-1}{|A|}} \nu(A) \\
        = & \ \sum\limits_{A \subseteq N, i \in A} \frac{1}{n \cdot \binom{n-1}{|A|-1}} \cdot \nu(A) - \sum\limits_{A \subseteq N, i \notin A} \frac{1}{n \cdot \binom{n-1}{|A|}} \nu(A) \\
        = & \ \sum\limits_{A \subseteq N, i \notin A} \frac{1}{n \cdot \binom{n-1}{|A|}} \cdot \left[ \nu(A \cup \{i\}) - \nu(A) \right] \\
        = & \ \phi_i \hfill \qed
	\end{array}
\end{equation*}

\newpage
\section{Analytical Solution and Sampling} \label{app:method}

\subsection{Analytical Solution to the Optimization Problem} \label{app:solution}

In order to solve the optimization problem presented in~\cref{eq:opt_kadd}, one may use a trick to remove the constraints.
One may include both $\emptyset$ and $N$, as well as $\nu(\emptyset)$ and $\nu(N)$, into the objective and assign them with large weights (e.g., $w_{\emptyset}=w_{N}=10^6$).
As a consequence, one ensures that both constraints
\begin{equation*}
    \nu(\emptyset) \ = \sum\limits_{\substack{B \subseteq N \\ |B| \leq k}} \gamma^{\left|B \right|}_0 I^k(B) \hspace{0.2cm} \text{and} \hspace{0.2cm} \nu(N) = \sum\limits_{\substack{B \subseteq N \\ |B| \leq k}} \gamma^{\left|B \right|}_{\left| B \right|} I^k(B)
\end{equation*}
 are satisfied when minimizing the objective which implies the constraint $\nu(N) - \nu(\emptyset) = \nu_k(N) - \nu_k(\emptyset)$ of the $k$-additive optimization problem.
With the aforementioned modifications, the optimization problem can be formulated as follows:
\begin{equation}
\label{eq:opt_kadd_mod}
\begin{array}{rl}
\displaystyle\min_{I^k} & \sum\limits_{A \in \mathcal{M}} w_{A}\left( \nu(A) - \sum\limits_{\substack{B \subseteq N \\ |B| \leq k}} \gamma^{\left|B \right|}_{\left| A \cap B \right|} I^k(B) \right)^2 \, .
\end{array}
\end{equation}
Clearly, (\ref{eq:opt_kadd_mod}) is a weighted least square problem.
Indeed, assume $\mathbf{W} \in \mathbb{R}^{T \times T}$ as a matrix whose diagonal elements are the weights $w_A$ for all sampled coalitions $A \in \mathcal{M}$, $| \mathcal{M} | = T$, $\nu_{\mathcal{M}}\in \mathbb{R}^{T \times 1}$ as the associated vector of sampled coalitions, and $\mathbf{P} \in \mathbb{R}^{T \times Q}$ as the transformation matrix from the generalized interaction indices to the game according to \cref{eq:iitomu_s} containing the weights $\gamma$. 
Further, let $I^k = (I^k(\emptyset), \phi_1^k, \ldots, \phi_n^k, I^k(\{1,2\}), \ldots, I^k(\{n-1,n\}), \ldots, I^k(A)) \in \mathbb{R}^{Q \times 1}$ with $\left| A \right| = k$ be the vector of generalized interactions in the lexicographic order for coalitions of players such that $\left| A \right| \leq k$.
For $k=n$, the interactions in $I^k$ can be set such that $\nu_{\mathcal{M}} = \mathbf{P} I^k$ as stated by \cref{eq:iitomu_s}.
However, as we are considering $k < n$, this optimization problem yields $I^k$ such that $\mathbf{P} I^k = \hat{\nu}_{\mathcal{M}}$ is as close as possible to $\nu_{\mathcal{M}}$. 
Note that $I^k$ is a $Q$-dimensional vector with $Q := \sum_{\ell=0}^k \binom{n}{\ell}$, where $n$ is the number of players.
In matrix notation, (\ref{eq:opt_kadd_mod}) can be formulated as
\begin{equation}
\label{eq:opt_kadd_ls}
\begin{array}{rl}
\displaystyle\min_{I^k} & \left( \nu_{\mathcal{M}} - \mathbf{P} I^k \right)^T \mathbf{W} \left( \nu_{\mathcal{M}} - \mathbf{P} I^k \right) \, ,
\end{array}
\end{equation}
whose well-known solution is given by
\begin{equation}
\label{eq:opt_kadd_ls2}
\begin{array}{rl}
I^k = \left( \mathbf{P}^T\mathbf{W}\mathbf{P} \right)^{-1} \mathbf{P}^T\mathbf{W} \nu_{\mathcal{M}} \, .
\end{array}
\end{equation}

\noindent
The asymptotic behavior of the computational complexity of solving (\ref{eq:opt_kadd_ls2}) involves:
\begin{itemize}
    \item Computing $\mathbf{P}^T\mathbf{W}\mathbf{P}$: matrix-matrix product yielding a complexity of $\mathcal{O} \left( TQ^2 \right)$;
    \item Computing $\mathbf{P}^T\mathbf{W} \nu_{\mathcal{M}}$: matrix-vector product yielding a complexity of $\mathcal{O} \left( TQ \right)$;
    \item Solving $\left( \mathbf{P}^T\mathbf{W}\mathbf{P} \right) I^k = \mathbf{P}^T\mathbf{W} \nu_{\mathcal{M}}$: yielding a complexity of $\mathcal{O} \left( Q^3 \right)$.
\end{itemize}
Therefore, as $\mathcal{O}\left( TQ \right)$ is dominated by the term $\mathcal{O} \left( TQ^2 \right)$, solving (\ref{eq:opt_kadd_ls2}) yields a complexity of $\mathcal{O} \left( TQ^2 + Q^3 \right)$.
In order to express this in terms of $n$ and $k$ we will bound $Q$ for $n \geq 2$ at the help of the inequality $\sum\limits_{\ell=1}^{k} \binom{n}{\ell} \leq n^k$ which we proof in the following.

\noindent \\
\textbf{Proof by induction over $k$:} \\
Induction base $k=1$:
\begin{equation*}
    \begin{array}{cc}
         \sum\limits_{\ell=1}^{1} \binom{n}{\ell} = \binom{n}{1} = n \leq n^1
    \end{array}
\end{equation*}
Induction hypothesis:
For arbitrary but fixed $n,k \in \mathbb{N}$ with $n \geq 2$ and $k \leq n$ holds: $\sum\limits_{\ell=1}^{k} \binom{n}{\ell} \leq n^k$. \\
Induction step $k \rightarrow k+1 \leq n$:
\begin{equation*}
    \begin{array}{cc}
         \sum\limits_{\ell=1}^{k+1} \binom{n}{\ell} = \binom{n}{k+1} + \sum\limits_{\ell=1}^{k} \binom{n}{\ell} \leq \frac{n!}{(k+1)! \cdot (n-k-1)!} + n^k \leq \frac{n^{k+1}}{2} + n^k = n^{k+1} \left( \frac{1}{2} + \frac{1}{n} \right) \leq n^{k+1}
    \end{array}
\end{equation*}
\qed

\noindent
Thus, we bound $Q$ by
\begin{equation*}
    \begin{array}{cc}
        Q = \sum\limits_{\ell=0}^k \binom{n}{\ell} = 1 + \sum\limits_{\ell=1}^k \binom{n}{\ell} \leq 1 + n^k \in \mathcal{O}(n^k)
    \end{array}
\end{equation*}
which results in a complexity polynomial in $n$ and linear in $T$ for solving the optimization problem:
\begin{equation}
    \mathcal{O} \left( n^{2k} T + n^{3k} \right) \, .
\end{equation}

\subsection{Sampling From Defined Probability Distribution} \label{app:sampling}

The pseudocode in \cref{alg:proposal} is simplified to ease understanding of the approximation procedure and thus it does not present the sampling of coalitions in a practical way.
Specifically, assigning a probability $p_A$ (see \cref{subsec:sampling}) used for sampling to each coalition within the powerset of $N$ causes exponential effort.
We outline in the following how to realize the sampling.

\noindent \\
Instead, one can exploit that the probabilities are equal for coalitions of the same size.
This allows to group the coalitions into strata with each stratum $\mathcal{S}_{\ell} := \{S \subseteq N \vert |S| = \ell \}$ containing all coalitions of size $\ell$.
Hence, all coalitions in $\mathcal{S}_{\ell}$ share the same probability $p_{\ell}$ equal to $p_A$ of any $A \in \mathcal{S}_{\ell}$ to begin with.
Thus, each stratum $\mathcal{S}_{\ell}$ has an initial weight, the sum of probabilities $\binom{n}{\ell} \cdot p_{\ell}$.
The weight of $\mathcal{S}_{\ell}$ can be updated accordingly by subtracting the probability $p_A$ (see line 8 in \cref{alg:proposal}) whenever a coalition $A \in \mathcal{S}_{\ell}$ is drawn.

\noindent \\
In each iteration one can now perform the sampling by
\begin{itemize}
    \item[1.] randomly selecting one of the $n-1$ many strata $\mathcal{S}_{1},\ldots,\mathcal{S}_{n-1}$ with probability proportional to its remaining weight,
    \item[2.] and uniformly at random drawing (without replacement) a remaining coalition $A$ of the chosen stratum $S_{\ell}$. 
\end{itemize}

\noindent
The second step is valid as each remaining $A \in \mathcal{S}_{\ell}$ has the same probability to be drawn.
Thus, this procedure simulates sampling (without replacement) according to the defined probabilities $p_A$ in \cref{subsec:sampling}.
\newpage
\section{Conceptual Comparison with KernelSHAP and $k_{ADD}$-SHAP}
\label{app:comparison}

The idea of using the Shapley values in (local or global) machine learning interpretability brings a difficulty in estimating the value function for all coalitions of features, specially when the number of features ($n$) is high. In \cite{Lundberg.2017}, the authors proposed the \textit{KernelSHAP}, an approach based on the Shapley value for local interpretability. The idea is to fit a locally weighted linear model $f(\cdot)$, by solving the optimization problem
\begin{equation}
\label{eq:kernelshap}
\begin{array}{rl}
\displaystyle\min_{\phi_0, \hat\phi_1, \ldots, \hat\phi_n} & \sum\limits_{z \in \mathcal{Z}} \pi_{z} \left( f(h(z)) - \phi_0 - \sum\limits_{i=1}^n z_i \hat\phi_i \right)^2 \, ,
\end{array}
\end{equation}
where
\begin{itemize}
    \item $\phi_0$ is the expected model output when no feature is present,
    \item $\hat\phi_1, \ldots, \hat\phi_n$ are the variables to be fitted, one for each feature, desirably yielding the Shapley values
    \item $z = \left[ z_1, \ldots, z_n \right]$ is a binary coalition vector such that $z_i = 1$ if feature $i$ is in the coalition, and 0 otherwise,
    \item $\mathcal{Z}$ is the set of all possible binary coalition vectors not containing only 0's and not containing only 1's,
    \item $\pi_{z} = \frac{n-1}{\binom{n}{|z|} |z| (n - |z|)}$, where $|z|$ indicates the number of features in the coalition,
    \item $h(z)$ is the simplified input.
\end{itemize}
The choice of $\pi_{z}$ is crucial in \textit{KernelSHAP}, as it ensures the convergence to the exact Shapley values as the number of coalitions in $\mathcal{M}$ increases based on a result in \cite{Charnes.1988}.
Worth noticing is the difference to our proposed method even for the simplest case of $k=1$.
Transferring our $k$-additive optimization problem from \cref{def:optimization} then yields the following objective function:
\begin{equation}
\label{eq:1additive}
\begin{array}{rl}
\displaystyle\min_{\phi_0, \hat\phi_1, \ldots, \hat\phi_n} & \sum\limits_{z \in \mathcal{Z}} \pi_{z}\left( f(h(z)) - \sum\limits_{i=1}^n (2\llbracket z_i = 1 \rrbracket -1) \phi_i \right)^2 \, ,
\end{array}
\end{equation}
where the variable $\hat\phi_i$ of a player $i$ not being contained in a coalition represented by $z$ now plays a role as it is subtracted from the inner sum, whereas for \textit{KernelSHAP} it has been left out due to $z_i=0$.

\noindent \\
In $k_{ADD}$-SHAP \cite{Pelegrina.2023}, instead of fitting a locally linear model, the authors proposed to fit the non-linear function called Choquet integral.
The interesting aspect in such a function is that its parameters are directly associated with the Shapley values ($\phi_1, \ldots, \phi_n$, represented by $I(\left\{ 1 \right\}), \ldots, I(\left\{ n \right\})$) as well as all (Shapley) interaction effects between coalitions of features ($I(B)$ for all $B \subseteq N$, where $N$ is the set of all features).
The proposed formulation leads to the following optimization problem:
\begin{equation}
\label{eq:kaddshap}
\begin{array}{rl}
\displaystyle\min_{I(B) \forall B \subseteq N} & \sum\limits_{A \in \mathcal{M}} \pi_{A}^{'}\left( f(h(z^A)) - \sum\limits_{B \subseteq N} \gamma^{\left|B \right|}_{\left| A \cap B \right|} I(B) \right)^2 \, ,
\end{array}
\end{equation}
where the weights $\pi_{A}^{'}$ have the same values for all $A$ except for the empty set and the grand coalition $N$ and $h(z^A)$ represents the simplified input such that $z_i^A = 1$ if $i \in A$, and 0 otherwise. In order to speed up the approximation of the Shapley values (and interaction effects), the authors proposed the use of $k$-additive games. If a game is assumed to be $k$-additive, all interaction effects with cardinality greater than $k$ is set to zero (\textit{i.e.}, $I(B) = 0$ for all $B$ such that $\left| B \right| > k$). By assuming $k$-additivity, one reduces the number of parameters while keeping enough flexibility to adjust the function even with lower $k$ (\textit{e.g.}, $k=2$ or $k=3$). This leads to the following optimization problem:
\begin{equation}
\label{eq:kaddshap2}
\begin{array}{rl}
\displaystyle\min_{I(B) \forall B \subseteq N, |B| \leq k} & \sum\limits_{A \in \mathcal{M}} \pi_{A}^{'}\left( f(h(z^A)) - \sum\limits_{\substack{B \subseteq N \\ |B| \leq k}} \gamma^{\left|B \right|}_{\left| A \cap B \right|} I(B) \right)^2 \, .
\end{array}
\end{equation}

\noindent
Note that both \textit{KernelSHAP} and $k_{ADD}$-SHAP method were conceived for local interpretability in machine learning. Surely, they can be extended to estimate global interpretability, for instance, by taking the average Shapley values over a set of local instances. Differently from these methods, in this paper, the proposed $SVAk_{\text{ADD}}$ can be used in any problem formulated as a game-theory scenario whose Shapley values must be estimated. As in \cite{Pelegrina.2023}, we also adopted the concept of $k$-additive games, however, we provide a general formulation to estimate the Shapley values. Moreover, we proved that by solving the proposed optimization problem
\begin{equation}
\begin{array}{rl}
\displaystyle\min_{I^k} & \sum\limits_{A \in \mathcal{M}} w_{A}\left( \nu(A) - \sum\limits_{\substack{B \subseteq N \\ |B| \leq k}} \gamma^{\left|B \right|}_{\left| A \cap B \right|} I^k(B) \right)^2 \, ,
\end{array}
\end{equation}
for the cases of $k=1$, $k=2$, and $k=3$, with weights $w_A^* = \binom{n-2}{|A|-1}^{-1}$, one obtains the Shapley values.
\newpage
\section{Cooperative Games Details} \label{app:cooperative_games}

The cooperative games used within our conducted experiments are based on explanation examples for real world data.
This section completes their brief description given in \cref{sec:exper} similar to \cite{Kolpaczki.2024b}.
Across all cooperative games the players represent a fixed set of features given by a particular dataset.

\subsection{Global Feature Importance}

Seeking to quantify each feature's individual importance to a model's predictive performance, the value function is based on the model's performance of a hold out test set.
This necessitates to split the dataset at hand into training and test set.
Features outside of an inspected coalition $S$ are removed by retraining the model on the training set and measuring its performance on the test set.
For all games we a applied train-test split of 70\% to 30\% and a random forest consisting of 20 trees.
For classification the value function maps each coalition to the model's resulting accuracy on the test set minus the accuracy of the mode within the data such that the empty coalition has a value of zero.
For regression tasks the worth of a coalition is the reduction of the model's mean squared error compared to the empty set which is given by the mean prediction.
Again, the empty coalition has a value of zero.

\subsection{Local Feature Attribution}

Instead of assessing each feature's contribution to the predictive performance, its influence on a model's prediction for a fixed datapoint can also be investigated.
Hence, the value function is based on the model's predicted value in comparison to the predicted value for the empty coalition.
For regression the model's predicted value is taken and for classification the assigned class probability of the predicted class in presence of the grand coalition.
Numerical features are removed by imputing their mean value and categorical features by imputing their mode.

\subsubsection{Adult Classification}
A sklearn gradient-boosted tree classifies whether a person's annual salary exceeds 50,000 in the \emph{Adult} \cite{Kohavi.1996} tabular dataset containing 14 features.
The predicted class probability of the true class is taken as the worth of a coalition $S$.
In order to render features outside of $S$ absent, these are imputed by their mean value such that the datapoint is compatible to the model's expected feature number.

\subsubsection{Image Classification}

A \emph{ResNet18} \cite{He.2016} model is used to classify images from \emph{ImageNet} \cite{Deng.2009}.
Since the for error tracking necessary exact computation of Shapley values is infeasible for the given number of pixels, 14 semantic segments are formed after applying \emph{SLIC} \cite{Achanta.2012}.
These super-pixels form the player set.
Given that the model predicts class $c$ using the full image, the value function assigns to each coalition $S$ the predicted class probability of $c$ resulting from only including those super-pixels in $S$.
The other super-pixels are removed by mean imputation, setting them grey.

\subsubsection{IMDB Sentiment Analysis}

A \emph{DistilBERT} \cite{Sanh.2019} transformer  fine-tuned on the \emph{IMDB} \cite{Maas.2011} dataset predicts the sentiment of a natural language sentence between -1 and 1.
The sentence is transformed into a sequence of tokens.
The input sentences are restricted to sentences that result in 14 tokens being represented by players of the cooperative game.
This allows to remove players in the tokenized representation of the transformer.
The predicted sentiment is taken as the worth of a coalition.

\subsection{Unsupervised Feature Importance}

In contrast to the previous settings, there is no available predictive model to investigate unlabeled data.
Still, each feature's contribution to the shared information within the data can be quantified and assigned as a score.
\citep{Balestra.2022} proposed to view the features $1,\ldots,n$ as random variables $X_1,\ldots,X_n$ such that the datapoints are realizations of their joint distribution.
Next, the worth of a coalition $S$ is given by their total correlation
\begin{equation*}
    \nu(S) = \sum\limits_{i \in S} H(X_i) - H(S)
\end{equation*}
where $H(X_i)$ denotes the Shannon entropy of $X_i$ and $H(S)$ the contained random variables joint Shannon entropy.
The utilized datasets are reduced in the number of features and datapoints to ease computation.
The \emph{Breast cancer} dataset contains 9 features and 286 datapoints.
The class label indicating the diagnosis is removed.
From the \emph{Big five} and \emph{FIFA 21} dataset 12 random features are selected out of the first 50 and the datapoints are reduced to the first 10,000.
This processing is carried out as done by \cite{Balestra.2022}.
\newpage
\section{Further Empirical Results}
\label{app:further_results}

\subsection{Impact of the Additivity Degree $k$}
\label{app:other_datasets}

In this section, we present the obtained results for other datasets. As mentioned in Section~\ref{subsec:degree}, the 1-additive model has less flexibility and, therefore, its error decreases more slowly in comparison to other additivity degrees for most datasets. Looking at the existing approaches, the proposed \emph{SVA}$k_{\text{ADD}}$ appears competitive, specially with \emph{Stratified SVARM}, and achieves similar performance with the \emph{KernelSHAP} for the Breast Cancer dataset.

\begin{figure*}[ht]
\centering
\begin{minipage}[c]{0.38\textwidth}
    \centering
    \includegraphics[width=0.99\textwidth]{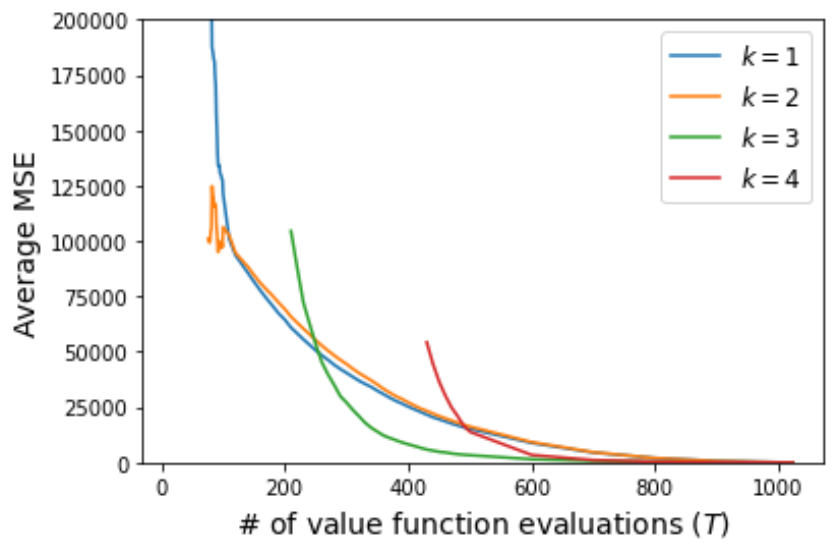}
    (a) Diabetes ($n=10$)
    \label{fig:diabetes_kadd}
\end{minipage}
\begin{minipage}[c]{0.38\textwidth}
    \centering
    \includegraphics[width=0.99\textwidth]{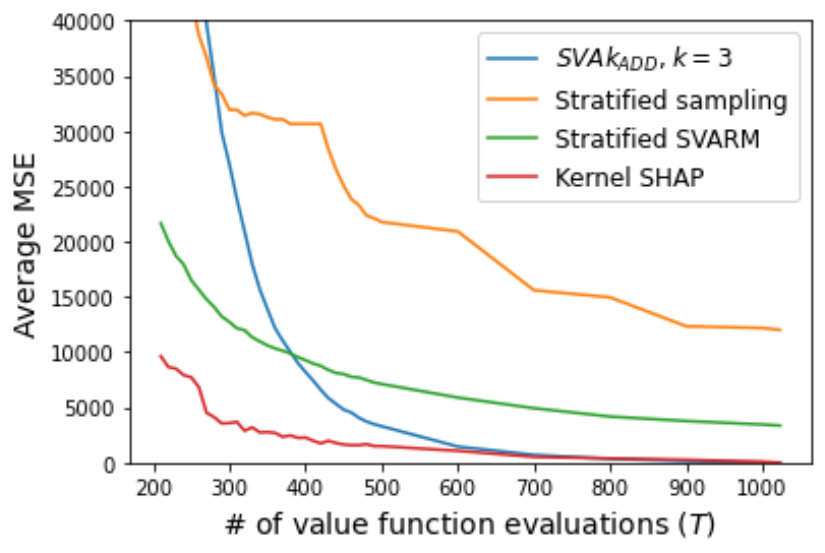}
    (b) Diabetes ($n=10$)
    \label{fig:diabetes_kadd_comp}
\end{minipage}
\begin{minipage}[c]{0.38\textwidth}
    \centering
    \includegraphics[width=0.99\textwidth]{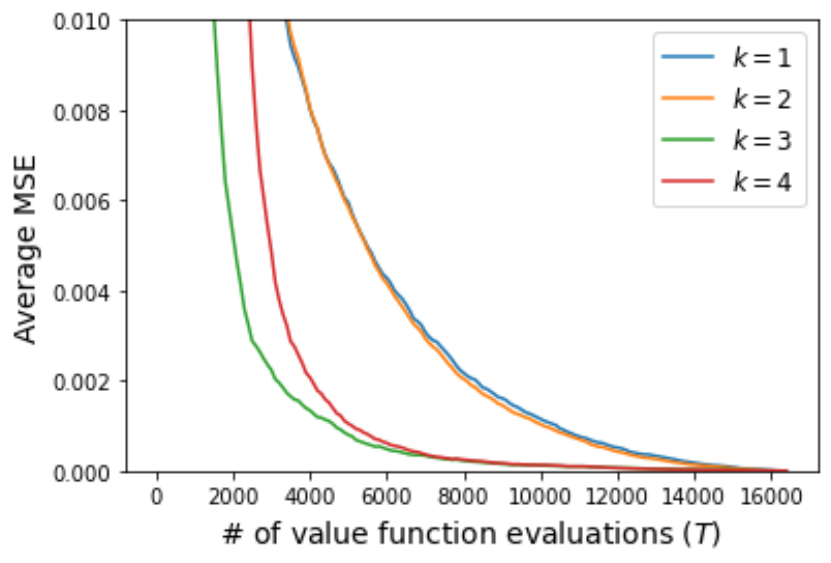}
    (c) IMDB ($n=14$)
    \label{fig:sentiment_kadd}
\end{minipage}
\begin{minipage}[c]{0.38\textwidth}
    \centering
    \includegraphics[width=0.99\textwidth]{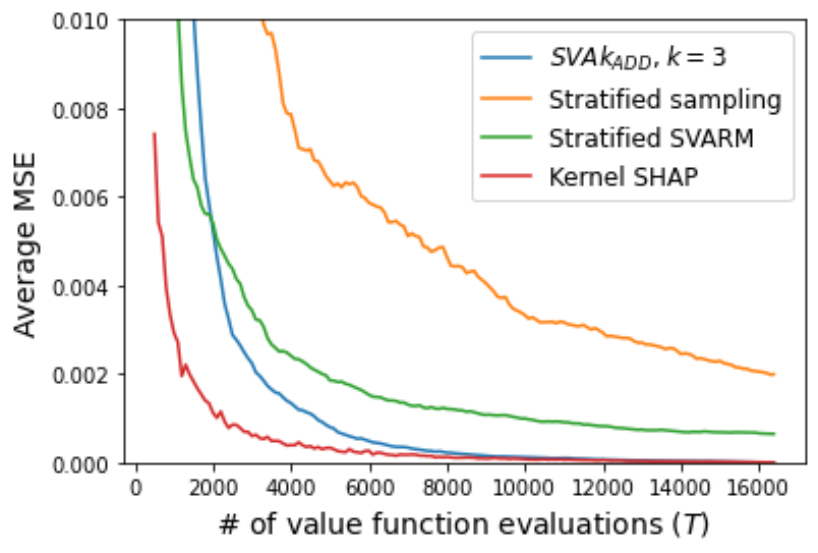}
    (d) IMDB ($n=14$)
    \label{fig:sentiment_kadd_comp}
\end{minipage}
\begin{minipage}[c]{0.38\textwidth}
    \centering
    \includegraphics[width=0.99\textwidth]{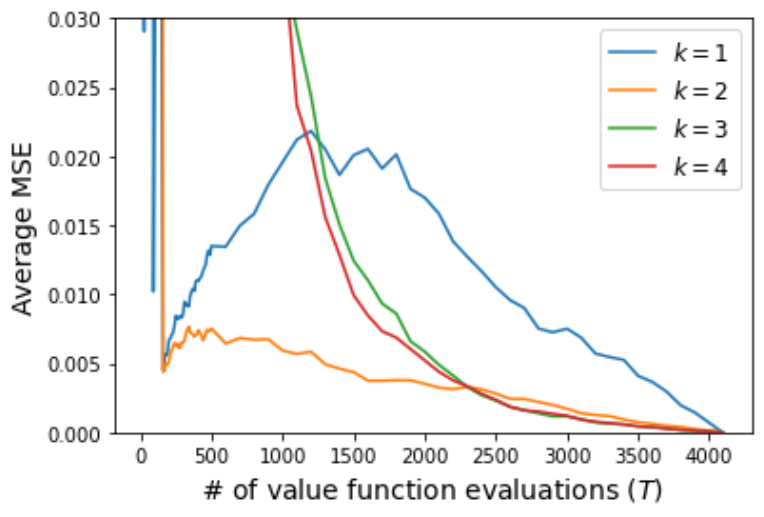}
    (e) FIFA 21 ($n=12$)
    \label{fig:fifa_kadd}
\end{minipage}
\begin{minipage}[c]{0.38\textwidth}
    \centering
   \includegraphics[width=0.99\textwidth]{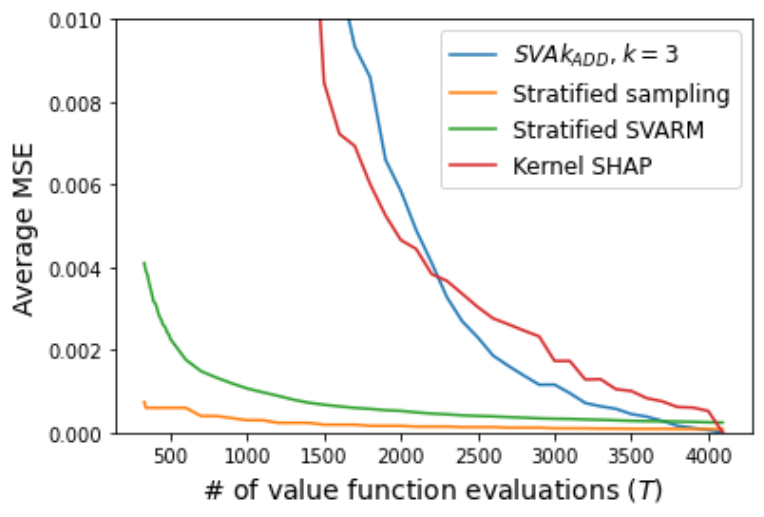}
    (f) FIFA 21 ($n=12$)
    \label{fig:fifa_kadd_comp}
\end{minipage}
\caption{MSE of \emph{SVA}$k_{\text{ADD}}$ and competing methods, averaged over 50 runs except for (c)-(f) with 10 runs, in dependence of available sample budget $T$. Datasets stem from various explanation types (i) global (first row), (ii) local (second row), and unsupervised (third row) with differing player numbers $n$.}
\label{fig:results_comp_other}
\end{figure*}

\newpage

\subsection{Comparison with Existing Approximation Methods} \label{app:futher_comparison}

\begin{figure*}[ht]
\centering
\begin{minipage}[c]{0.32\textwidth}
    \centering
    \includegraphics[width=0.99\textwidth]{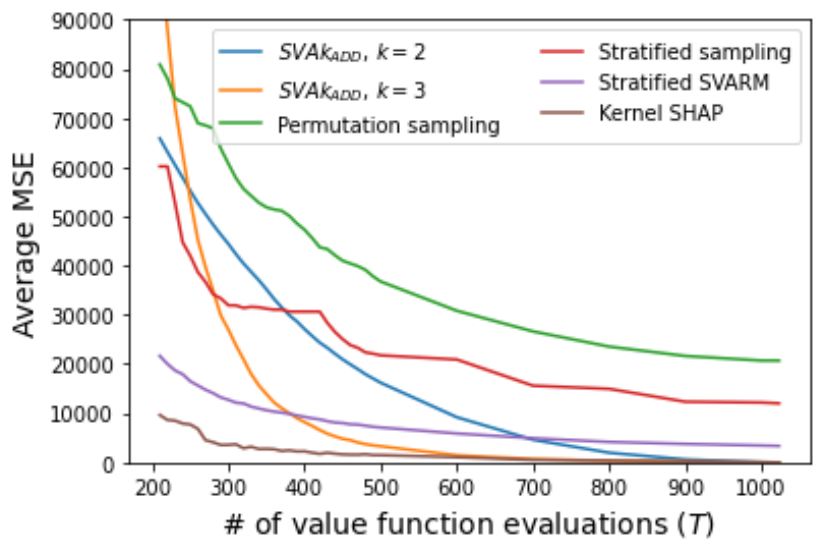}
    (a) Diabetes ($n=10$)
    \label{fig:diabetes_kadd_comp_all}
\end{minipage}
\begin{minipage}[c]{0.32\textwidth}
    \centering
    \includegraphics[width=0.99\textwidth]{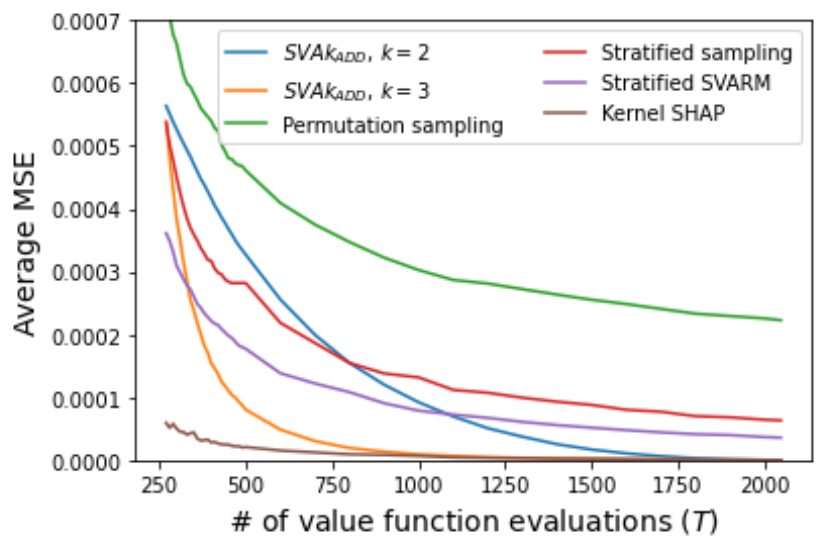}
    (b) Titanic ($n=11$)
    \label{fig:titanic_kadd_comp_all}
\end{minipage}
\begin{minipage}[c]{0.32\textwidth}
    \centering
    \includegraphics[width=0.99\textwidth]{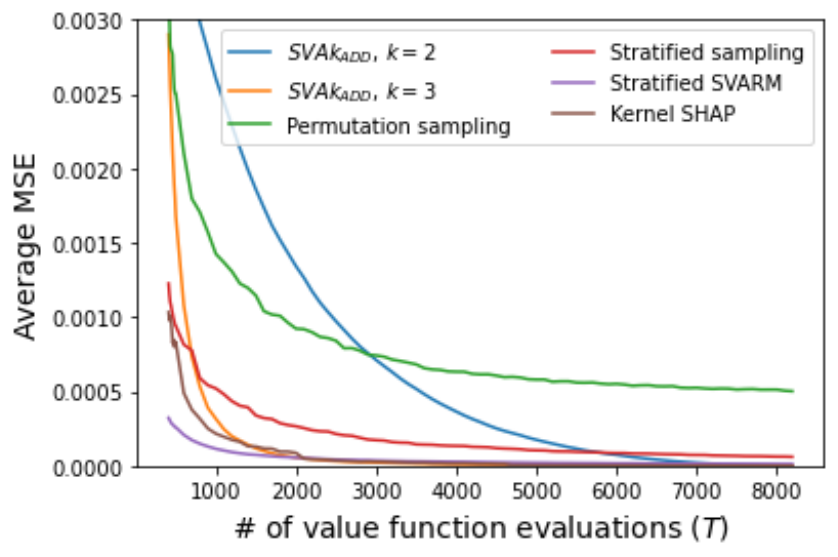}
    (c) Wine ($n=13$)
    \label{fig:wine_kadd_comp_all}
\end{minipage}
\begin{minipage}[c]{0.32\textwidth}
    \centering
    \includegraphics[width=0.99\textwidth]{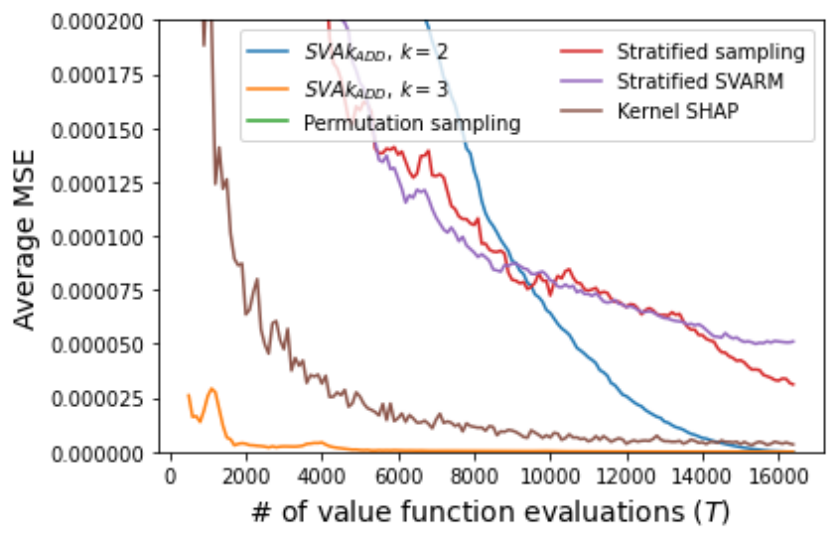}
    (d) Adult ($n=14$)
    \label{fig:adult_local_kadd_comp_all}
\end{minipage}
\begin{minipage}[c]{0.32\textwidth}
    \centering
    \includegraphics[width=0.99\textwidth]{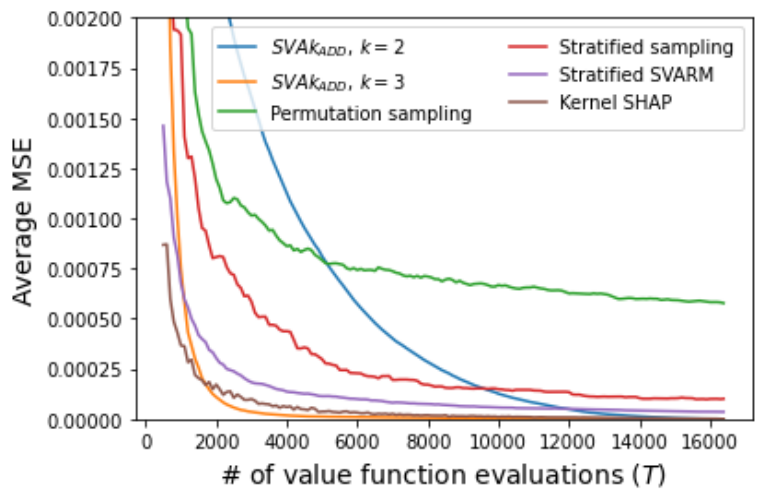}
    (e) ImageNet ($n=14$)
    \label{fig:image_cat_kadd_comp_all}
\end{minipage}
\begin{minipage}[c]{0.32\textwidth}
    \centering
    \includegraphics[width=0.99\textwidth]{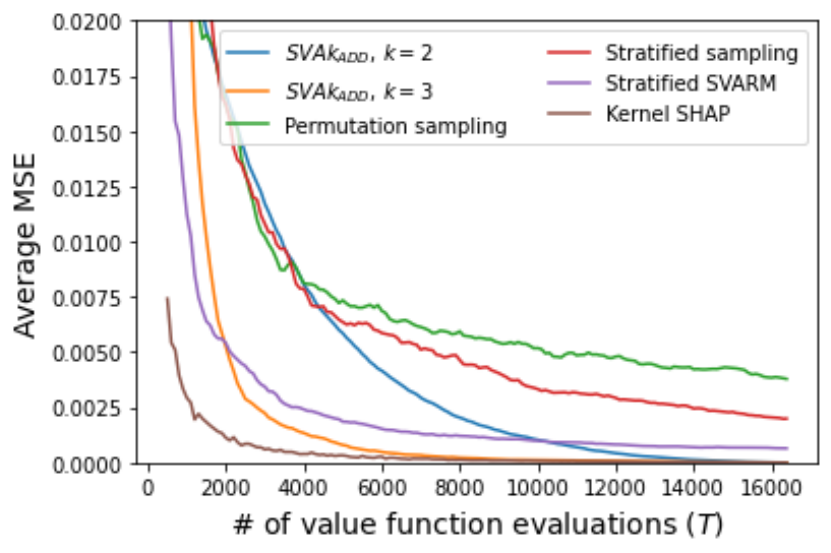}
    (f) IMDB ($n=14$)
    \label{fig:sentiment_kadd_comp_all}
\end{minipage}
\begin{minipage}[c]{0.32\textwidth}
    \centering
    \includegraphics[width=0.99\textwidth]{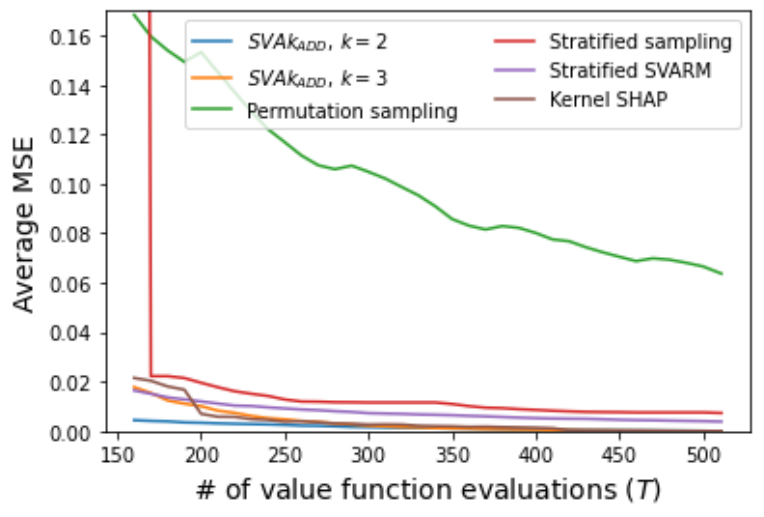}
    (g) Breast Cancer ($n=9$)
    \label{fig:breastcancer_kadd_comp_all}
\end{minipage}
\begin{minipage}[c]{0.32\textwidth}
    \centering
    \includegraphics[width=0.99\textwidth]{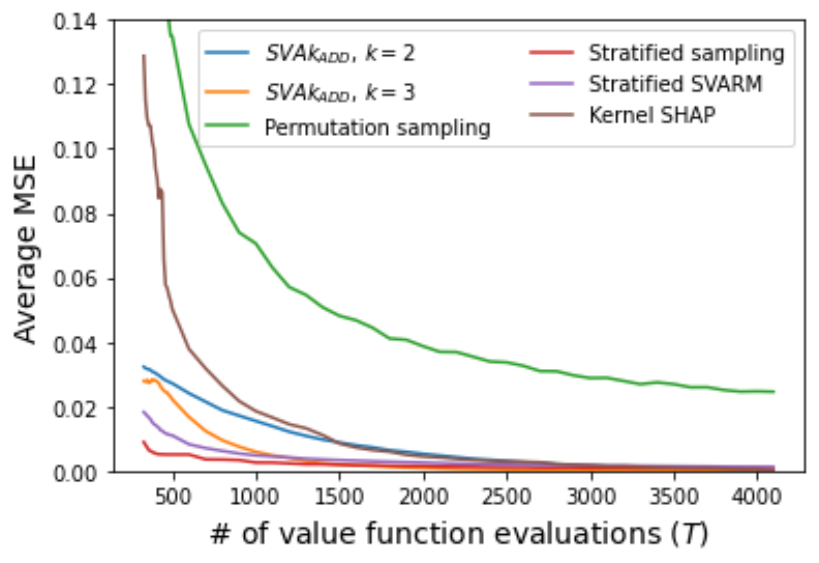}
    (h) Big Five ($n=12$)
    \label{fig:bigfive_kadd_comp_all}
\end{minipage}
\begin{minipage}[c]{0.32\textwidth}
    \centering
    \includegraphics[width=0.99\textwidth]{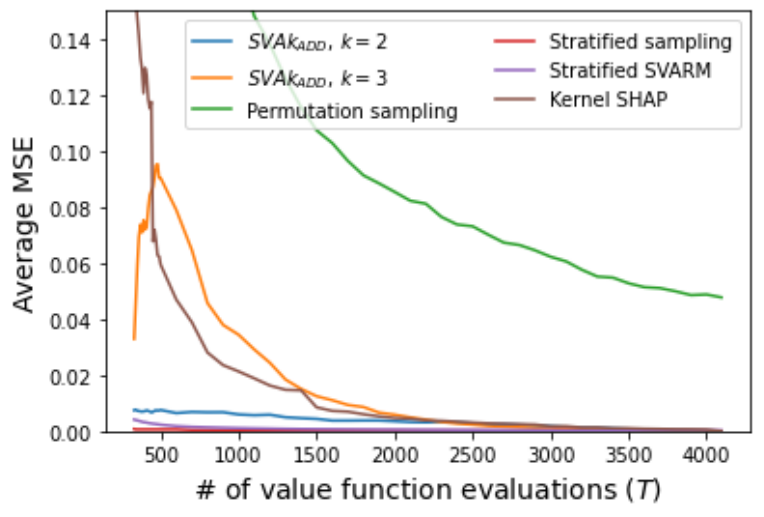}
    (i) FIFA 21 ($n=12$)
    \label{fig:fifa_kadd_comp_all}
\end{minipage}
\caption{MSE of \emph{SVA}$k_{\text{ADD}}$ and competing methods, averaged over 50 runs except for (d), (e), (f) and (i) with 10 runs, in dependence of available sample budget $T$. Datasets stem from various explanation types (i) global (first row), (ii) local (second row), and unsupervised (third row) with differing player numbers $n$.}
\label{fig:results_comp_all}
\end{figure*}

Aiming at further illustrating the performance of our proposal in comparison with existing approximation methods, in this section, we extend the results presented in Section~\ref{subsec:performance} by including another approach, called \emph{ApproShapley} (given here as \emph{Permutation sampling}) \cite{Castro.2009}, and our proposal for $k=2$. We show this comparison in Figure~\ref{fig:results_comp_all}. Note that, with the exception of the Wine dataset, the \emph{Permutation sampling} leads to the worst results. 

\end{document}